\documentclass[print]{elsarticle}
\journal{Pattern Recognition}

\usepackage{amssymb}
\usepackage{subfigure}
\usepackage{algorithm}
\usepackage{algorithmic}
\renewcommand{\algorithmicrequire}{\textbf{Input:}}  
\renewcommand{\algorithmicensure}{\textbf{Output:}}  
\usepackage{multirow}
\usepackage[flushleft]{threeparttable}
\usepackage{rotating}
\usepackage{float}
\usepackage{amsthm}
\usepackage{geometry}
\usepackage{amsmath}
\usepackage{color}
\usepackage{setspace}

\begin{document}

\begin{frontmatter}

\title{Cross-domain Network Representations}


\author[mymainaddress]{Shan Xue}
\ead{shan.xue@student.uts.edu.au}

\author[mymainaddress]{Jie Lu\corref{mycorrespondingauthor}}
\cortext[mycorrespondingauthor]{Corresponding author}
\ead{jie.lu@uts.edu.au}

\author[mymainaddress]{Guangquan Zhang}
\ead{guangquan.zhang@uts.edu.au}

\address[mymainaddress]{Center for Artificial Intelligence, Faculty of Engineering and Information Technology, University of Technology Sydney, Australia}

\begin{abstract}
The purpose of network representation is to learn a set of latent features by obtaining community information from network structures to provide knowledge for machine learning tasks. Recent research has driven significant progress in network representation by employing random walks as the network sampling strategy. Nevertheless, existing approaches rely on domain-specifically rich community structures and fail in the network that lack topological information in its own domain. In this paper, we propose a novel algorithm for cross-domain network representation, named as CDNR. By generating the random walks from a structural rich domain and transferring the knowledge on the random walks across domains, it enables a network representation for the structural scarce domain as well. To be specific, CDNR is realized by a cross-domain two-layer node-scale balance algorithm and a cross-domain two-layer knowledge transfer algorithm in the framework of cross-domain two-layer random walk learning. Experiments on various real-world datasets demonstrate the effectiveness of CDNR for universal networks in an unsupervised way.
\end{abstract}


\end{frontmatter}


\section{Introduction}
\label{sec:introduction}

Networks generated from mature systems usually have larger numbers of entities such as nodes and edges than the emerging ones. For example, a new born online social media attracts limited numbers of users and hasn't formed massive interactions among them, from where the it gets extremely scarcer scale than the mature media like Facebook. Furthermore, in some domains such as the biological domain, it's difficult to collect sufficient data due to the costs, technique barriers, ethic reasons and so on. Traditional industries normally lack historical data when data-driven techniques haven't brought them benefits. Above scenarios lead to data deficiency which affect network analysis and learning. Previous approaches developed for network representation based on large-scale datasets are not able to be applied. 

\begin{figure}[!b]
	\small
	\centering
	\includegraphics[width=0.7\linewidth]{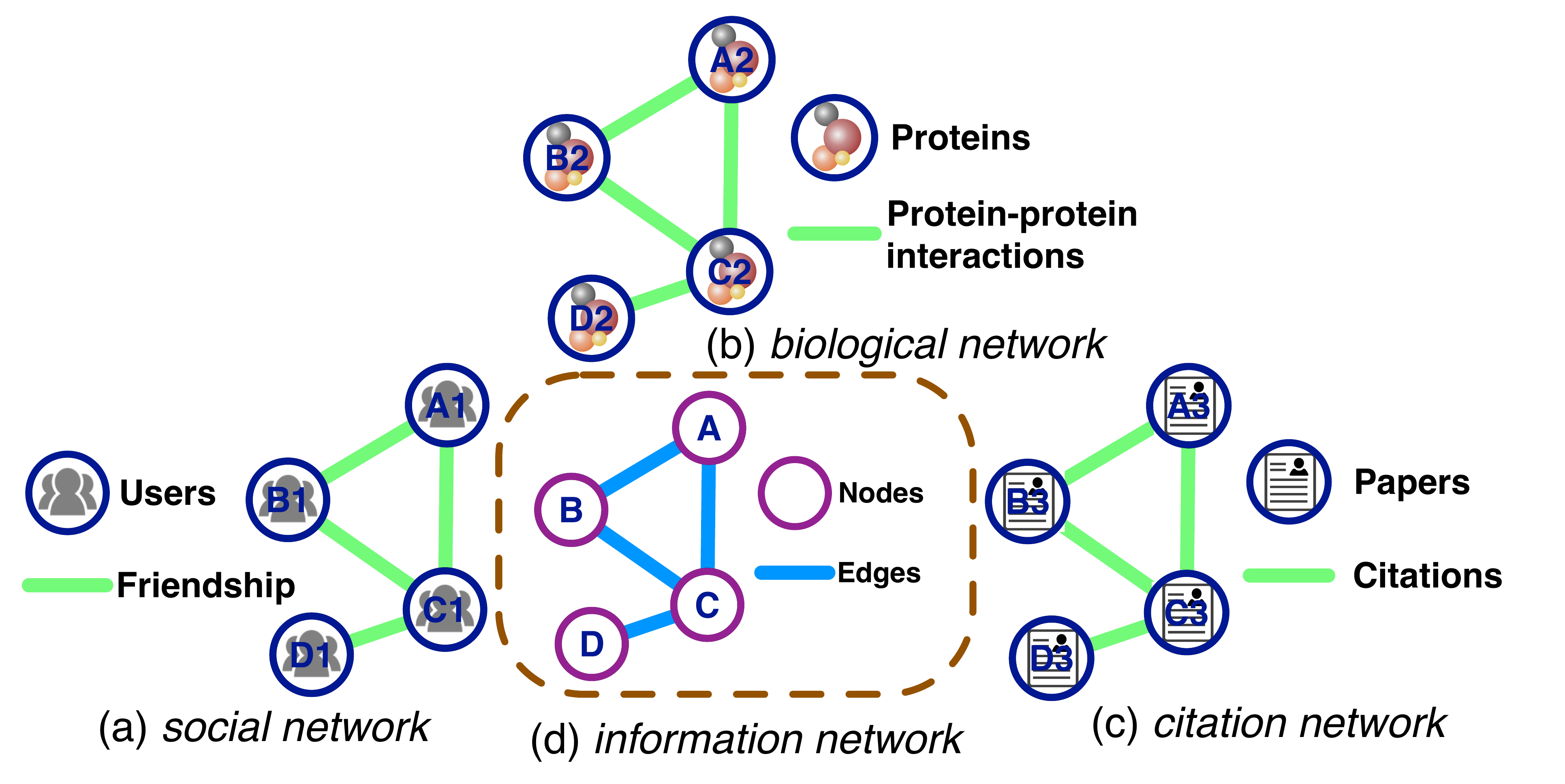}
	\caption{Illustrations of undirected network structures formed by entities of nodes and edges. (a) Social network is formed by users \{A1,B1,C1,D1\} and user friendships; (b) Biological network is formed by proteins \{A2,B2,C2,D2\} and protein-protein interactions; and (c) Citation network is formed by papers \{A3,B3,C3,D3\} and citations. (d) Information network extracts information flows from (a), (b) and (c) as edges and inherit the nodes \{A,B,C,D\}.}
	\label{fig:NetStrucBig}
\end{figure}

From the domain-specific view, rich data collected from real-world complex systems with large-scale network datasets. The components in a system are defined as the nodes in a network, direct interactions between nodes are defined as edges, and connection strengths are described by weights on edges. Techniques not only analysis networks but also learn knowledge from network structures which has become a main stream in network research for artificial intelligence purposes \cite{xuan2016uncertainty, wang2018learning}. To this end, networks are preliminarily categorized based on real-world systems and their physical properties, such as social network \cite{chaker2017social,choi2012incremental}, biological network \cite{jancura2010dividing} and citation network \cite{lu2018structural}. As shown in Figure \ref{fig:NetStrucBig}, social networks (a) denote users as nodes and friendship as edges; biological networks such as the Protein-Protein Interactions (PPI) network (b) models proteins as nodes and PPI as edges; and citation networks (c) represent papers as nodes and citations as edges. 

From all kinds of networks, the information network \cite{tang2015line} abstract the information flows from the original network structure, where the original nodes like users, proteins and authors are treated as the information users and suppliers and the information interchanges on friendship, PPI and citations as edges. The information network encode the network behaviors and save them into the network structure as shown in Figure \ref{fig:NetStrucBig}(d). Information networks help us illustrate the entities in a physical system but raise a question on how to understand the various properties behind the different network categories especially when the topologies seem no difference as shown in Figure \ref{fig:NetStrucBig}.

Network representation aims to learn a latent feature/vector space by learning from the information formed by network entities \cite{tang2009relational}. It inputs the high-dimensional network structures and outputs relatively low-dimensional representations in encoding as many community properties as possible. For the use of machine learning, network representation should output complex but highly structured latent features, to meet the smoothness requirement in learning function and to overcome the sparsity from input data \cite{bengio2013representation}. To this end, a series of network representation approaches have been proposed based on the sampling strategy of random walks and the deep learning technique in the last decade. The random walk is a type of similarity measurement for a variety of problems in community detection \cite{noh2004random,lai2010enhanced}, which computes the local community structure information sub-linear to the size of the input network \cite{yang2015defining,leskovec2012learning}. A stream of short random walks is used as a basic tool for extracting information from real-world large-scale information networks \cite{liu2010link, perozzi2014deepwalk}.

The typical random walk-based network representation algorithms, such as DeepWalk \cite{perozzi2014deepwalk}, learn sequences of nodes as a stream of short random walks to model the network structures of deep features which obviously are highly dependent on the sliced window that controls random walk learning for node sampling purpose. However, when the distance between two nodes is larger than the sliced window size, the random walk jumps to the next round. Although it could be covered by introducing a vast amount of sampling, the repetitions increase computational complexity. This explains the main reason why the networks with small structure scales are barely applicable for these algorithms. Therefore, the previous works on random walk-based network representations are limited in a domain-specific way so that the performance mainly relies on the network topological quality. Our previous work proposed a framework for transferring structures across large-scale information networks (FTLSIN) \cite{IJCNN18}, however only enabled structural knowledge transfer across relational information networks and both networks should have large scales. The cases listed in the beginning of this paper will not be guaranteed satisfying latent feature spaces from the limited network structures for the further machine learning tasks within one domain.

To address above problems, we propose a novel algorithm for universal cross-domain network representations (CDNR) with the following contributions. 

\begin{enumerate}
	\item[1)] CDNR offers an effective learning solution for the network representation, where the network doesn't have enough entities that causes a random walk failure in structural sampling. 
	
	\item[2)] CDNR determines the relationships between two independent networks which would belong to irrelevant domains. Similar network patterns are detected so that links generated between the corresponding communities transfer knowledge in CDNR.
	
	\item[3)] CDNR predicts the potential entities for the scarce network structures by employing the cross-domain two-layer random walk (\textit{CD2L-RandomWalk}) framework from \cite{IJCNN18} and integrating two novel algorithms, cross-domain two-layer node-scale balance (\textit{CD2L-NodeBalance}) and cross-domain two-layer knowledge transfer (\textit{CD2L-KnowlTransfer}).
\end{enumerate}

The rest of the paper is arranged as follows. In Section \ref{sec:relatedworks}, related works are summarized. In Section \ref{sec:problem}, we state the CDNR problem. In Section \ref{sec:CD2L}, the proposed CDNR algorithm is explained in detail. In Section \ref{sec:experiment}, two experiments are designed to evaluate the representations on real-world datasets. Our conclusions are presented in Section \ref{sec:Conclusions}.

\section{Related Works}
\label{sec:relatedworks}

The previously used per-node partition function \cite{bandyopadhyay2008counting} is expensive to compute, especially for large information networks. To overcome this disadvantage, a series of sampling strategies have been proposed \cite{kurant2011towards,lelis2013predicting} to analyze the statistics within local structures, e.g., communities and sub-networks. These approaches are different from traditional representation learning \cite{paalanen2006feature,zhu2015unsupervised,ding2015deep}. The latent feature learning of the network representation captures neighborhood similarity and community membership in topologies \cite{yang2015network,pan2016tri,tu2016max}. 

DeepWalk \cite{perozzi2014deepwalk} trains a neural language model on the random walks generated by the network structure. After denoting a random walk that starts from a root node, DeepWalk slides a window and maps the central node to its representation. Hierarchical Softmax factors out the probability distributions corresponding to the random walk and the representation function is updated to maximize the probability. DeepWalk has produced promising results in dealing with sparsity in scalable networks, but has relatively high computational complexity for large-scale information networks. LINE, Node2Vec and Struc2Vec are the other structure-based network representation algorithms that improve the performance of DeepWalk. LINE \cite{tang2015line} preserves both the local network structure and the global network structure by first-order proximity and second-order proximity respectively and can be applied to large-scale deep network structures that are directed, undirected, weighted and unweighted. Node2Vec \cite{grover2016node2vec} explores the diverse neighborhoods of nodes in a biased random walk procedure by employing classic search strategies. Struc2Vec \cite{ribeiro2017struc2vec} encodes structural similarities and generates the structural context for nodes using random walks. The above-mentioned works has contributed to network analysis by modeling a stream of short random walks. 

\begin{figure}[!htp]
	\small
	\centering
	\subfigure[dblp]{\includegraphics[width=0.23\linewidth]{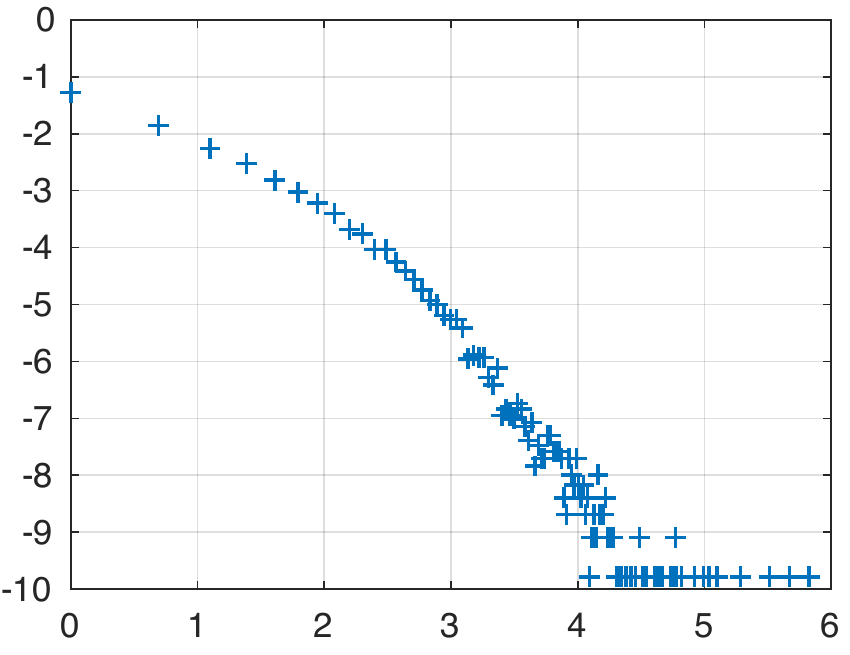}}
	\subfigure[dblp RW]{\includegraphics[width=0.23\linewidth]{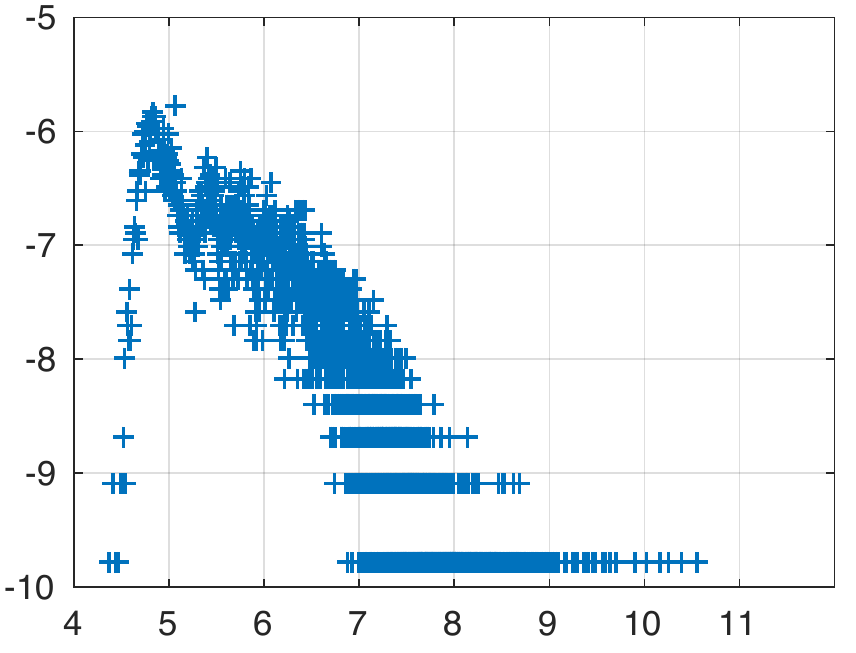}}
	\subfigure[M10]{\includegraphics[width=0.23\linewidth]{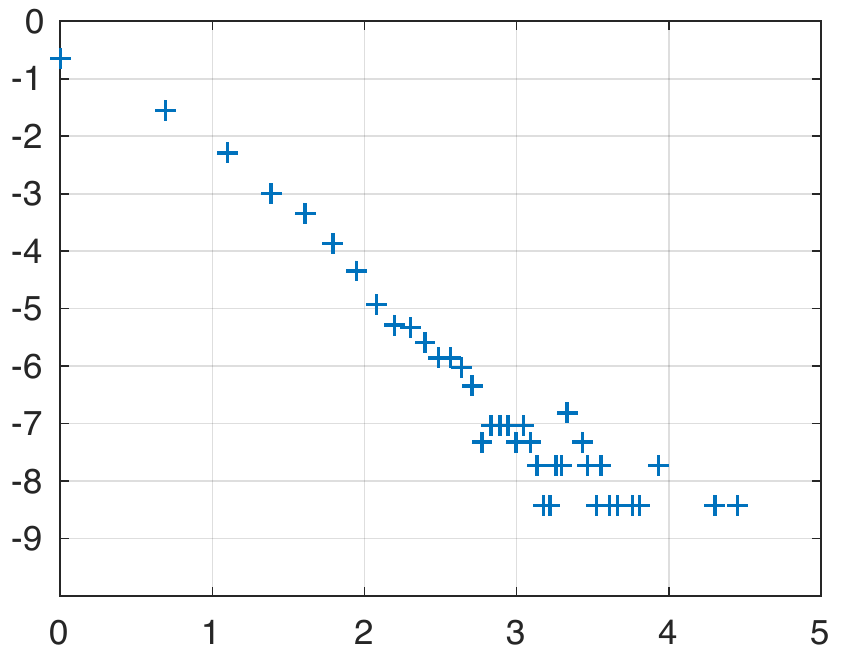}}
	\subfigure[M10 RW]{\includegraphics[width=0.23\linewidth]{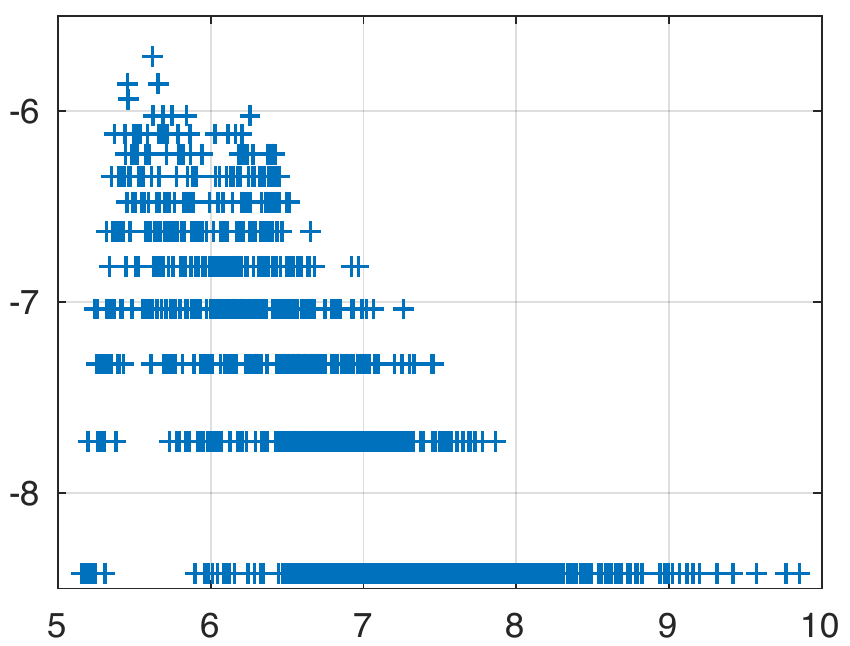}}
	\caption{Power-law distribution of the single-label classification datasets and their random walks on the networks. The X-axial is denoted as $\log(Deg)$ of the network and the Y-axial is denoted as $\log(Pr(Deg))$. Each power-law distribution pair of the network and its random walks should follow the same pattern so that random walks over the network can conduct skip-gram based network representations. For example, (a) and (b) are formed as a power-law distribution pair following the same pattern by which random walks on the dblp network are guaranteed a network representation on dblp.}
	\label{fig:powerlawsingle}
\end{figure}

All the previous works based on random walk to sample networks into a steam of nodes are under a common assumption of power-law distribution. The power-law distribution exists widely in real-world networks. It is a special degree distribution that follows $P(deg)\sim deg^{-a}$, where $deg$ is a node degree and $a$ is a positive constant \cite{newman2005power}. A network that follows the power-law distribution is also regarded as a scale-free network with the scale invariance property \cite{barabasi2009scale}. The social networks, biological networks and citation networks being discussed in this paper are observed to be scale-free in nature \cite{barabasi2016network}. In $log$-$log$ axes, the power-law distribution shows a linear trend on the slope ratio of $-a$ (Figure \ref{fig:powerlawsingle} and Figure \ref{fig:powerlawmulti}), which reflects that numerous edges connect small degree nodes and will not change regardless of network scale \cite{adamic2000power}. It has been observed in \cite{perozzi2014deepwalk} that if a network follows the power-law distribution, the frequency at which a node undertakes in a short random walk will also follow the same distribution. Meanwhile, random walks in power-law distribution networks naturally gravitate towards high degree nodes \cite{adamic2001search}. 

In this paper, we propose CDNR which employs biased random walk sampling strategies to learn network structures based on previous works. However, CDNR is different from the deep transfer learning approaches for cross-domain graph-structured data, i.e., context enhanced inductive representation \cite{hamilton2017inductive}, intrinsic geometric information transfer \cite{lee2017transfer} and deep inductive graph representation \cite{rossi2018deep}. Deep neural network-based network representation usually need to generalize a small set of base feature for deep learning, such as network statistical properties like node degree, which lost valuable information from networks. The link predictions in \textit{CD2L-RandomWalk} are therefore leveraged on the power-law distribution as well as the distance calculation between the two independent networks across domains. The network that has small distance to the target network is regarded as the source domain. The scale invariance property should theoretically ensure that power law-based CDNR is robust. 

\section{Problem Statement}
\label{sec:problem}

\begin{table}[!b]
	\centering \small
	\renewcommand\arraystretch{0.6}
	\caption{Summary of notations.}
	\label{tab:notations}
	\begin{tabular}{rl}
		\hline
		$\mathcal{D}^s$, $\mathcal{D}^t$ & A source domain and a target domain. \\
		$\textbf{X}^t$, $\mathbf{x}^t$, $x^t_i$ & A $d$-dimensional target domain latent feature spaces, a $N^t$-dimensional feature \\
		 & vectors, and the $i$-th element of $\mathbf{x}^t$. \\
		$G^s$, $G^t$ & A (un)directed unattributed unweighted network from $\mathcal{D}^s$, and a (un)directed \\
		& unattributed weighted network from $\mathcal{D}^t$. \\
		$V^s$, $V^t$, $v^s$, $v^t$ & The node set of $G^s$, the node set of $G^t$, a node in $V^s$, and a node in $V^t$. \\
		$E^s$, $E^t$, $e^s_{ij}$, $e^t_{ij}$ & The edge set of $G^s$, the edge set of $G^t$, an edge between $v^s_i$ and $v^s_j$, and an edge \\
		 & between $v^t_i$ and $v^t_j$. \\
		$W^t$, $w^t_{ij}$ & The weight set on $E^t$, and a weight in $W^t$ on $e^t_{ij}$. \\
		$\mathcal{G}$, $\mathcal{V}$, $E'$ & A form of super graph for $G^s$, a super-node set, and a super-edge set. \\
		$V'$, $e'_{V'_iV'_j}$ & A super node in $\mathcal{V}$, and a super edge in $E'$ connecting $V'_i$ and $V'_j$. \\
		$W'$, $w'_{V'_iV'_j}$ & A weight set, and a weight in $W'$ on $e'_{V'_iV'_j}$. \\
		$\mathcal{W}^s$, $\mathcal{W}^t$ & The random walk sets on $G^s$ and $G^t$. \\
		$\mathcal{P}_{V'_iV'_j}$ & A shortest path between $V'_i$ and $V'_j$ over $\mathcal{G}$. \\
		$deg^s$, $deg^t$, $deg'$ & A set of node degree values on $G^s$, a set of node degree values on $G^t$ and a set of \\
		& node degree values on $\mathcal{G}$. \\
		$E^*$, $W^*$ & A link set between $G^t$ and $\mathcal{G}$ across domains, and a weight set on $E^*$.\\
		$e^*_{v^tV'}$, $w^*_{v^tV'}$ & A link in $E^*$ that connects $v^t$ and $V'$, and a weight in $W^*$ on $e^*_{v^tV'}$. \\
		$a$ & The slope ratio of power-law distribution. \\
		$Deg(\cdot)$ & The function calculates the node degree. \\
		$\langle \cdot \rangle$ & The average function on a value set. \\
		$|\cdot|$ & The function counts the number in a set. \\
		\hline
	\end{tabular}
\end{table}

\textbf{Definition 1 (Domain \cite{lu2015transfer})} A domain is denoted as $\mathcal{D}=\{\textbf{X}, P(\mathbf{x})\}$, where $\textbf{X}$ is the feature space and $P(\mathbf{x})$ is the marginal probability distribution that $\mathbf{x}=\{x_1,\cdots,x_n\}\in \textbf{X}$. 
	
\textbf{Definition 2 (Network \cite{barabasi2016network})} Let $G=(V,E,W)$ be a given network, where $V$ represents one kind of entities known as nodes, $E$ represents another kind of entities known as edges reflecting connections between nodes, $E\subseteq (V\times V)$, and $W$ represents the possible weights on $E$.

\textbf{Definition 3 (Cross-domain Network Representation)} Suppose a source domain $\mathcal{D}^s=\{\textbf{X}^s,P(\mathbf{x}^s)\}$ represented by $G^s=(V^s,E^s)$ and a target domain $\mathcal{D}^t=\{\textbf{X}^t,P(\mathbf{x}^t)\}$ represented by $G^t=(V^t,E^t,W^t)$, domains are irrelevant $\mathcal{D}^s \neq \mathcal{D}^t$ if $\textbf{X}^s \neq \textbf{X}^t$ and $P(\mathbf{x}^s) \neq P(\mathbf{x}^t)$; and relevant if $\textbf{X}^s = \textbf{X}^t$ or $P(\mathbf{x}^s) = P(\mathbf{x}^t)$. CDNR employs structural information on $W^t$, from $G^s=(V^s,E^s)$ to $G^t=(V^t,E^t,W^t)$, to improve the target domain representations $f\colon V^t\rightarrow \textbf{X}^t$ in a $d$-dimensional latent feature space.

To prepare the structural knowledge from the source domain, CDNR firstly implements a maximum likelihood optimization to generate a set of random walks $\mathcal{W}^s$ on $G^s$ in the bottom layer of \textit{CD2L-RandomWalk}. A neighborhood $N_S(u^s)\subset G^s$ is clustered rooted at node $u^s$ by the neighborhood sampling strategy $S$ on the biased random walks \cite{grover2016node2vec}. Then,  \textit{CD2L-RandomWalk} constructs links between $G^s$ and $G^t$. To this end, \textit{CD2L-NodeBalance} balances the scales of $V^s$ and $V^t$ by clustering $V^s$ into super nodes $V' \in \mathcal{V}$ in which $v^s\in V'$ share close node degrees with $v^t \in V^t$; and generate links between $V^t$ and $\mathcal{V}$. \textit{CD2L-KnowlTransfer} trains the maximized similarities across two domains and determines how much value should be transferred across the shortest paths $\mathcal{P}$ and $E^t$, where $\mathcal{P}$ are formed by the super edges $E'$ and the values are save in $W^t$. In CDNR, the representations $\textbf{X}^t$ are learned in the top layer of \textit{CD2L-RandomWalk} and will be evaluated by a standard classification task.

\subsection{Bottom-layer Random Walk: Knowledge Preparation}

The bottom-layer random walk is designed for knowledge preparation in the source domain. The sampled random walks contains structural knowledge from which will be transferred to the target domain. The bottom-layer random walk introduces a biased random walk strategy to efficiently explore diverse neighborhoods and sample the nodes along the shortest path\footnote{The shortest path is a path between two nodes for which the sum of its edge weights is minimized.}. Suppose a set of random walks $\mathcal{W}^s$, each root node $v^s$ repeats $k$ times for sampling and each random walk is set in a length of $l$. In generating a random walk, suppose we are standing at node $c$ which is the $i$-th node in the random walk, $1<i<l$, the node $c-1$ denotes the $i-1$-th node and the $i+1$-th node $x$ is chosen from $N_S(c)$ based on a probability $P(x|c)=\frac{\pi_{xc}}{Z}$ where $Z$ is the partition function that ensures a normalized distribution \cite{bengio2013representation} and $\pi_{xc}=\alpha_{pq}(x,c)$ is guided by the search bias $\alpha_{pq}$. To be specific, $\alpha_{pq}(x,c)$ follows the searching rules: if the length of the shortest path between nodes $x$ and $c-1$ is $|\mathcal{P}_{xc-1}|=0$, then $\alpha_{pq}(x,c)=1/p$; $\alpha_{pq}(x,c)=1$, if $|\mathcal{P}_{xc-1}|=1$; and $\alpha_{pq}(x,c)=1/q$, if $|\mathcal{P}_{xc-1}|=2$. The sampling strategy on the biased random walks is computationally efficient especially for real-world large-scale networks.

\section{Knowledge Transfer in Cross-domain Network Representations}
\label{sec:CD2L}

CDNR enables the cross-domain random walk-based network representations and assumes both networks across domains follow the power-law distribution. Representations in CDNR work under the Skip-gram framework and are optimized by maximum likelihood over biased random walks. The contributions of CDNR are realized in this section by \textit{CD2L-RandomWalk} with the two components: \textit{CD2L-NodeBalance} and \textit{CD2L-KnowlTransfer}.

\subsection{Cross-domain Two-layer Node-scale Balance and Link Prediction}

By transferring knowledge from an external source domain, CDNR deals with the scenarios that the training sample in the target domain is insufficient to make a good network representation. Such knowledge transfer belongs to a transfer learning task \cite{pan2010survey} arises two questions: 1) Link prediction: how to construct paths between two networks across domains for \textit{CD2L-RandomWalk}, and 2) \textit{CD2L-NodeBalance}: how to solve the problem of unbalanced node scales.

The unbalancedness between $G^s$ and $G^t$ is reflected on the nodes $|V^s|>|V^t|$ and also on the connections $\langle deg^s \rangle>\langle deg^t \rangle$, where $|V^s|$ and $|V^t|$ refer to the node scales, and $\langle deg^s \rangle$ and $\langle deg^t \rangle$ refer to the average node degrees\footnote{The average degree is a mean on the degrees of all nodes in the network.}. In this case, \textit{CD2L-NodeBalance} tries to reform $G^s$ into a smaller size based on the network structures of $G^t$. For the purpose of discovering sub-graph patterns \cite{wang2017incremental}, a concept of super node \cite{guo2014super} is employed and we define the formation for CDNR.

\textbf{Definition 4 (Super Node in Source Domain)}
A super node is a sub-graph of the original source network. Denoting the super-node set $\mathcal{V}$, a super node $V' \in \mathcal{V}$ consists of a group of nodes $\{v^s\}\subseteq V^s$ and the edges $\{e^s\}\subseteq E^s$ connecting to or from $\{v^s\}$. The nodes $\{v^s\}$ that clustered into a $V'$ have close node degrees.

To cluster a set of nodes in the large-scale network, a super-node learning based on the nodes in the target domain is as follows:

\begin{equation}\label{eq:fnode}
\Phi_{Snode}\colon V^s \rightarrow \{V',e^*_{v^tV'},w^*_{v^tV'}|V^t\}
\end{equation}
where $e^*_{v^tV'}$ is a predicted link between $v^t\in V^t$ and $V'=\{v^s\}$ across domains, and $w^*_{v^tV'}$ is the weight on $e^*_{v^tV'}$ which indicates the similarity between $v^t$ and $V'$ and how much knowledge should be transferred from the source domain to the target domain in \textit{CD2L-RandomWalk}. 

\textit{CD2L-NodeBalance} attempts to pair each node $v^t$ with at least one super node $V'$ in a minimum super-node scale $|\mathcal{V}|$ and a maximum likelihood between $V^t$ and $\mathcal{V}$ according to Eq. (\ref{eq:nodemappingopt}). For each pair of $(v^t,V')$, we firstly initialize a link and a weight following,

\begin{equation}\label{eq:cde}
e^*_{v^tV'}=\left\{
\begin{array}{ll}
1 & \text{if}~w^*_{v^tV'}>0\\
0 & \text{if}~w^*_{v^tV'}=0
\end{array}
\right.
\end{equation}

\begin{equation}\label{eq:cdw}
w^{*(0)}_{v^tV'}=\frac{\min(Deg(v^t),Deg(V'))}{\max(Deg(v^t),Deg(V'))}
\end{equation}
where $Deg(v^t)$ denotes the degree of $v^t$, $Deg(V')$ denotes the degree of $V'$, and $V'$ is initialized on nodes in the same degree.

To optimize $\Phi_{Snode}$ in Eq. (\ref{eq:nodemappingopt}), we analysis the degree ranges over $V^t$ and $V'$ in $[1,\max(deg^t)]$ and $[1,\max(deg')]$ respectively and reorganize $V'$ including merging and dividing super nodes based on the following three cases.

\begin{figure}[!b]
	\small
	\centering
	\includegraphics[width=0.7\linewidth]{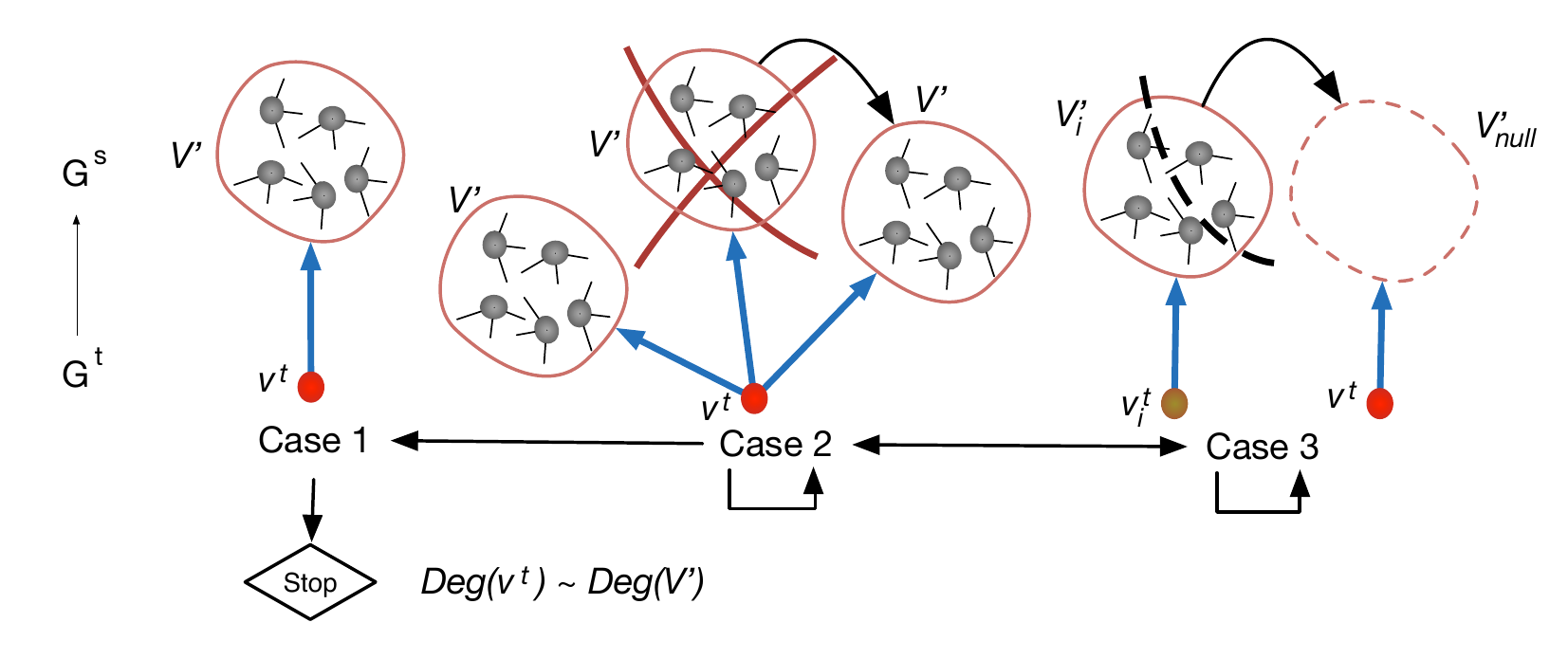}
	\caption{An illustration of the three cases in \textit{CD2L-NodeBalance} and their interconversions: 1) Case 2 to Case 1, 2) Case 2 to Case 2, 3) Case 2 to Case 3, 4) Case 3 to Case 2 and 5) Case 3 to Case 3.}
	\label{fig:nodebalance}
\end{figure}

Denoting the range scales $n_{deg^t}=|deg^t|$ and $n_{deg'}=|deg'|$, there are three possible cases of \textit{CD2L-NodeBalance} as follows and as shown in Figure \ref{fig:nodebalance}. Degree sets $deg^t$ and $deg'$ are always ranked in a decreasing order. A $v^t$ finds the corresponding $V'$ that are in the same position in $deg^t$ and $deg'$, denoted as $Deg(v^t)\sim Deg(V')$.

Case 1: If $n_{deg^s}=n_{deg^t}$, only one $V'$ links to $v^t$. In this case, \textit{CD2L-NodeBalance} is completed in the initialization stage with $E^*=\{e^*_{v^tV'}\}$ and $W^*=\{w^*_{v^tV'}\}$.

Case 2: If $n_{deg^s}>n_{deg^t}$, more than one $V'$ links to $v^t$. $W^*=\{w^*_{v^tV'}\}$ at the current stage is going to be optimized in Eq. (\ref{eq:nodemappingopt}). If $w^*_{v^tV'}$ turns to 0, the edge $e^*_{v^tV'}$ is deleted and the $V'$ is merged into another super node that linked with $v^t$ and gets the smallest weight.

Case 3: If $n_{deg^s}<n_{deg^t}$, there are at least one $v^t$ not linked to any $V'$. We add a group of empty super nodes $V'_{null}$ in a number of $n_{deg^t}-n_{deg^s}$ and evenly insert them into $\mathcal{V}$. To fill up the $V'_{null}$, a few nodes in $V'\neq \emptyset$ next to $V'_{null}$ are removed and added to $V'_{null}$. $V'_{null}$ then is initialized $w^{*(0)}_{v^tV'}=0$ by Eq. (\ref{eq:cdw}).

In Case 2 and Case 3, $\Phi_{Snode}$ is optimized by maximizing the likelihood between $V^t$ and $\mathcal{V}$. Starting from each $v^t_i$, a vector $\vec{w}=[w_1,\cdots,w_i,\cdots,w_{n_{node}}]^{\top}$ weights for each pair of $(v^t_i,V'_j)$, where $i=1,\cdots, |V^t|$, $j=1,\cdots,n_{link}$ and $n_{link}=\max(n_{deg'},n_{deg^t})$. If there is link between $(v^t_i,V'_j)$, $w_i=w^*_{v^tV'}$; else wise 0.

\begin{equation}\label{eq:nodemappingopt}
\max_{\Phi_{Snode}}\sum_i{\eta \sum_j{\big\lbrack\log{(C)}-a_{\oplus}\log{(\vec{\delta}_z \vec{w} \vec{w}^{\top} \vec{\delta}_z^{\top})}\big \rbrack}}
\end{equation}
where $\vec{\delta}_z$ is a vector in size of $n_{link}$ with the value of 0 or 1, which based on $Deg(v^t)\sim Deg(V')$ in Cases 1-3. Let $a_{\oplus}=\min\{a^s,a^t\}$ in which $a^s$ and $a^t$ are the power-law slope ratio of $G^s$ and $G^t$ respectively. $\eta=\frac{1}{n_{deg^t}}e^{\frac{1-{n^2_{deg'}}}{n_{deg'}}}\gamma e^\lambda$ controls the range of the likelihood over \textit{CD2L-NodeBalance}, where $\gamma$ is a parameter for $V^t$ and $\lambda$ is a parameter for $\mathcal{V}$. The optimized \textit{CD2L-NodeBalance} results suggest the predicted links $E^*\propto \vec{\delta}_z$ where $W^*=\vec{\delta}_z \vec{w}$ in Case 2 and Case 3.

\begin{algorithm}[!t]
	\small
	\setstretch{0.8}
	\caption{The \textit{CD2L-NodeBalance} algorithm.}\label{alg:nodebalance}
	\renewcommand{\algorithmicrequire}{\textbf{Input:}}
	\renewcommand\algorithmicensure {\textbf{Initialize:}}
	\begin{algorithmic}[1]
		\REQUIRE ~~\\
		$G^t=(V^t,E^t)$ in the target domain and $G^s=(V^s,E^s)$ in the source domain.
		\ENSURE ~~ \\
		$V'^{(0)} \leftarrow$ Cluster $v^s \in V^s$ by node degrees. \\
		$n^{(0)}_{deg^s}$, $n^{(0)}_{deg^t}\leftarrow$ Node degree scales in $G^s$ and $G^t$.\\
		$W^{*(0)} \leftarrow$ Apply Eq. (\ref{eq:cdw}). \\
		\WHILE{$n^{(t)}_{deg^s} \neq n^{(t)}_{deg^t}$}
		\STATE $W^{*(t)}\leftarrow$ Apply Eq. (\ref{eq:nodemappingopt})
		\STATE $E^{*(t)}\leftarrow$ Apply Eq. (\ref{eq:cde})
		\ENDWHILE {~$\{\epsilon=\max^{(t)}_{\Phi_{Snode}}-\max^{(t-1)}_{\Phi_{Snode}}\}$}
		\RETURN $E^*=E^{*(t)}$ and $W^*=W^{*(t)}$
	\end{algorithmic}
\end{algorithm}

\subsection{Cross-domain Two-layer Knowledge Transfer and Target Domain Edge Evolvement}

\begin{algorithm}[!b]
	\small
	\setstretch{0.8}
	\caption{The \textit{CD2L-KnowlTransfer} algorithm.}\label{alg:knowltransfer}
	\renewcommand{\algorithmicrequire}{\textbf{Input:}}
	\renewcommand\algorithmicensure {\textbf{Initialize:}}
	\begin{algorithmic}[1]
		\REQUIRE ~~\\
		$\mathcal{W}^s$ Random walks of $G^s$ generated in the bottom-layer of \textit{CD2L-RandomWalk};
		$G^t=(V^t,E^t,W^{t(0)})$ in the target domain;
		and $E^*$ and $W^*$ from Algorithm \ref{alg:nodebalance}. \\
		\FOR{$e^t_{ij}$ in $E^t$}
		\STATE $w'_{V'_iV'_j} \leftarrow$ Apply Eq. (\ref{eq:weightonsuperedge}).
		\STATE $\mathcal{P}_{V'_iV'_j} \leftarrow$ Construct shortest paths between $V'_i$ and $V'_j$.
		\STATE $w^t_{ij} \leftarrow$ Update weight on $e^t_{ij}$ by Eq. (\ref{eq:WMWall}).
		\STATE $e^t_{ij} \leftarrow$ Evolve new edge if $w^{t(0)}_{ij}=0$ and $w^t_{ij}>0$.
		\ENDFOR
		\RETURN $G^t=(V^t,E^t,W^t)$
	\end{algorithmic}
\end{algorithm}

\begin{figure}[!t]
	\small
	\centering
	\includegraphics[width=0.6\linewidth]{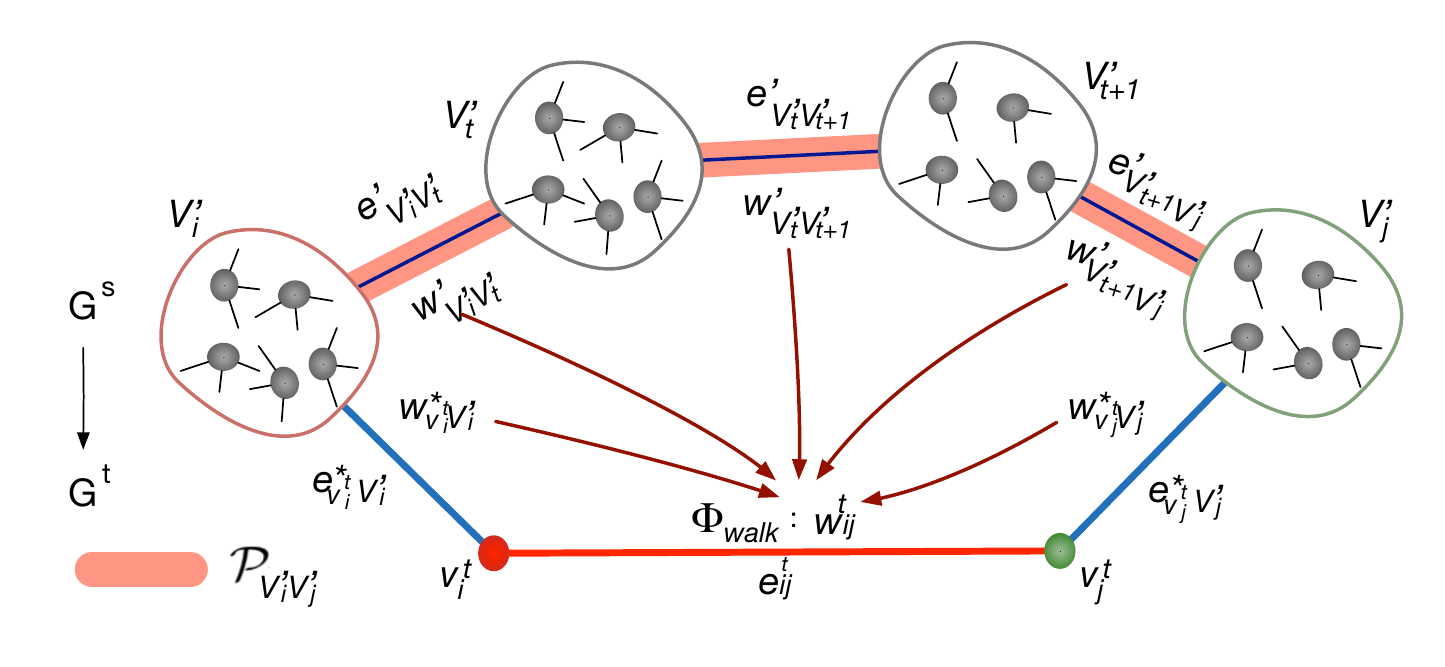}
	\caption{An illustration of weight contributions on a target domain network edge in \textit{CD2L-KnowlTransfer}.}
	\label{fig:knowltransfer}
\end{figure}

\textit{CD2L-KnowlTransfer} transfers the knowledge saved in weights through the predicted links $E^*$. The knowledge includes three parts of weights as shown in Figure \ref{fig:knowltransfer}: a weight on the super edge that reflects the knowledge learning from the random walks in the source domain, two weights on the predicted links, and the original weight on $e^t$ (in this paper is 1 or 0). The \textit{CD2L-KnowlTransfer} follows:

\begin{equation}\label{eq:walkmapping}
\Phi_{Knowl}\colon (G^t,\mathcal{G},W^*)\rightarrow W^t
\end{equation}
where $\mathcal{G}$ denotes the super graph.

\textbf{Definition 5 (Super Graph in Source Domain)} A super graph $\mathcal{G}=(\mathcal{V},E',W')$ reformed from $G^s$ is formed by super nodes $\mathcal{V}=\{V'\}$, super edges $E'=\{e'_{V'_iV'_j}\}$ and the super weights $W'=\{w'_{V'_iV'_j}\}$ on $E'$, where $\mathcal{F}\colon (\mathcal{W}^s,\mathcal{V})\rightarrow E'$. If a random walk belongs to $\mathcal{W}^s$ goes through $V'_i$ and $V'_j$, there will be an $e'_{V'_iV'_j}$. 

\begin{equation}\label{eq:weightonsuperedge}
	w'_{V'_iV'_j}=\sum_{v^s_i\in V'_i} \sum_{v^s_j\in V'_j} \frac{1}{d_{\mathcal{W}^s}(v^s_i, v^s_j)}
\end{equation}
where $d_{\mathcal{W}^s}(v^s_i, v^s_j)$ is the distance between nodes $v^s_i$ and $v^s_j$ in a random walk, and $w'_{V'_iV'_j}$ calculates every random walk going over $e'_{V'_iV'_j}$.

In having the three parts of weights $\{w'_{V'_iV'_j}, w^*_{v^t_iV'_i}, w^*_{v^t_jV'_j}, w^{t(0)}_{ij}\}$ that contribute to $w^t_{ij}$ in \textit{CD2L-KnowlTransfer}, the weight on $e^t_{ij}\in E^t$ in the top layer of the \textit{CD2L-RandomWalk} are denoted as:

\begin{equation}\label{eq:WMWall}
w^t_{ij} = w^{t(0)}_{ij} + \frac{1}{Z} \sum_{V'_i} \sum_{V'_j} {w^*_{v^t_iV'_i} \cdot w^*_{v^t_jV'_j} \cdot \big\lbrack \frac{1}{l_\mathcal{P}}\sum_{e'_{V'_tV'_{t+1}}\subseteq \mathcal{P}_{V'_iV'_j}} w'_{V'_tV'_{t+1}}\big\rbrack}
\end{equation}
where $w^{t(0)}_{ij}=\{0,1\}$ is the original weight that reflects an edge between $v^t_i$ and $v^t_j$ or no edge, $Z$ is for normalization, $\mathcal{P}_{V'_iV'_j}$ is the shortest path between $V'_i$ and $V'_j$ over $\mathcal{G}$, $l_\mathcal{P}$ is the length of $\mathcal{P}_{V'_iV'_j}$, $e'_{V'_tV'_{t+1}}\in E'$ denotes an edge that consists in $\mathcal{P}_{V'_iV'_j}$, and $w'_{V'_tV'_{t+1}}$ is the weight on $e'_{V'_tV'_{t+1}}$.

In \textit{CD2L-KnowlTransfer} above, $G^t$ is enriched in network structures by putting extra weights on the original edges and also evolves possible edges. 

\subsection{Top-layer Random Walk and Network Representations}

\begin{algorithm}[!t]
	\small
	\caption{The CDNR algorithm.}\label{alg:CDNR}
	\renewcommand{\algorithmicrequire}{\textbf{\textit{Top-layer Feature Learning}}}
	\renewcommand\algorithmicensure {\textbf{\textit{CD2L-RandomWalk}}}
	\begin{algorithmic}[1]
		\ENSURE ~~\\
		\STATE $\mathcal{W}^s \leftarrow$ Random walks generated from $G^s$ in the Bottom-layer Random Walk.
		\STATE $E^*$, $W^* \leftarrow$ Apply Algorithm \ref{alg:nodebalance}.
		\STATE $W^t \leftarrow$ Apply Algorithm \ref{alg:knowltransfer}.
	\end{algorithmic}
	\begin{algorithmic}[1]
		\REQUIRE ~~\\
		\FOR{$u^t$ in $G^t$}
		\STATE $N_S(u^t) \leftarrow$ Search neighborhood of $u^t$ with $W^t$.
		\STATE $f \leftarrow$ Apply Skip-gram to optimize.
		\ENDFOR
		\RETURN $\mathbf{X}^t\leftarrow$ A latent feature space of $G^t$ by $f$.
	\end{algorithmic}
\end{algorithm}

CDNR represents $G^t$ in the top layer of \textit{CD2L-RandomWalk} after \textit{CD2L-NodeBalance} and \textit{CD2L-KnowlTransfer}. CDNR learns the latent feature space by $f\colon V^t \rightarrow \mathbf{X}^t$ in the Skip-gram framework.

Given a node $u^t$ in the target domain with the window size $r$, we obtain a cross-domain Skip-gram for $G^t$ by maximizing the following log-likelihood function of $f$ in observing a neighborhood of $N_S(u^t)$,

\begin{equation}\label{eq:target_skip-gram}
\max_{f}~~\sum_{u^t\in V^t}\log{Pr(N_S(u^t)|f(u^t))}
\end{equation}
where $\mathcal{W}^t$ is learned on $P(x^t|u^t)=\frac{\pi_{x^tu^t}}{Z}$ that $\pi_{x^tu^t}=\alpha_{pq}(x^t,u^t)\cdot w^t_{x^tu^t}$, while $\mathcal{W}^s$ is learned on $\pi_{x^su^s}=\alpha_{pq}(x^s,u^s)$.

In summary, Algorithm \ref{alg:CDNR} of CDNR is formed by \textit{CD2L-RandomWalk} and \textit{Top-layer Feature Learning}. The main advantage of CDNR is that when the network representation is poor because it lacks structures, the \textit{CD2L-RandomWalk} enables knowledge transfer from external domains and CDNR doesn't need to rebuild a network representation model. CDNR offers an efficient cross-domain learning with a relatively low computational cost of $O(\langle deg^t \rangle |V^t|)$ on \textit{CD2L-NodeBalance}, $O(|E^t|)$ on \textit{CD2L-KnowlTransfer} and $O({\langle deg^t \rangle}^2|V^t|)$ on network representation in line with Node2Vec \cite{grover2016node2vec}.

\section{Experiments}
\label{sec:experiment}

This section evaluates the effectiveness of the CDNR compared to the baseline algorithms of network representations in both single-label classifications (Section \ref{subsec:singlelabel}) and multi-label classifications (Section \ref{subsec:multilabel}).

\subsection{Baseline Algorithms}

This experiment evaluates the performance of the unsupervised CDNR on the target networks. The representation outputs are applied to a standard supervised learning task, i.e., linear SVM classification \cite{suykens1999least}. The experiments choose a simple classifier because we want to put less emphasis on classifiers in evaluate the network representation performance. The baseline algorithms are chosen from the previous domain-specific network representations and a deep inductive graph representation as follows.

\begin{enumerate}
	
	\item[$\bullet$] \textbf{DeepWalk} (Perozzi \textit{et al.} 2014) \cite{perozzi2014deepwalk} is the first random walk-based network representation algorithm. By choosing DeepWalks, we exclude the matrix factorization approaches which have already been demonstrated to be inferior to DeepWalk.
	
	\item[$\bullet$] \textbf{LINE} (Tang \textit{et al.} 2015) \cite{tang2015line} learns latent feature representations from large-scale information networks by an edge-sampling strategy in two separate phases of first- and second-order proximities. We excluded a recent Graph Factorization algorithm \cite{ahmed2013distributed} because LINE demonstrated better performance in the previous experiment.
	
	\item[$\bullet$] \textbf{Node2Vec} (Grover \textit{et al.} 2016) \cite{grover2016node2vec} learns continuous feature representations of nodes using a biased random walk procedure to capture the diversity of connectivity patterns observed in networks with the biased parameter $\alpha$ which is controlled by parameters of $p$ and $q$.
	
	\item[$\bullet$] \textbf{Struc2Vec} (Ribeiro \textit{et al.} 2017) \cite{ribeiro2017struc2vec} learns node representations from structural identity by constructing a hierarchical graph to encode structural similarities and generating a structural context for nodes.
	
	\item[$\bullet$] \textbf{DeepGL} (Rossi \textit{et al.} 2018) \cite{rossi2018deep} learns interpretable inductive graph representations by relational functions for each representing feature and achieve inductive transfer learning across networks. It inputs a 3-dimensional base features to a CNN and outputs the representation in $d$ dimensions where $d$ depends on learning.
	
\end{enumerate}

\subsection{Experiment on Single-label Dataset}\label{subsec:singlelabel}

\subsubsection{Single-label Datasets}

Two academic citation networks are selected as the datasets. Both of them are used for the multi-class classification problem \cite{galar2011overview}. Nodes are denoted as papers in these networks.

\begin{table}[!h]
	\small
	\centering
	\addtolength{\tabcolsep}{7pt}
	\caption{Single-label classification dataset statistics.}
	\label{tab:SingleLabelDataset}
	\begin{tabular}{ccccc}
		\hline
		\multirow{2}*{\textbf{Domain}} & \multirow{2}*{\textbf{Datasets}} & \textbf{Num. of} & \textbf{Num. of} & \textbf{Num. of}\\
		& & \textbf{Nodes} & \textbf{Edges} & \textbf{Categories}\\
		\hline
		Source & DBLP & 60,744 & 52,890 & 4\\
		Target & M10 & 10,310 & 77,218 & 10\\
		\hline
	\end{tabular}
\end{table}

\begin{enumerate}
	
	\item[$\bullet$] \textbf{DBLP dataset\footnote{http://arnetminer.org/citation (V4 version is used)}} (source network) consists of bibliographic data in computer science. Each paper may cite or be cited by other papers, naturally forming a citation network. The network in this dataset abstracts a list of conferences from four research areas, \textit{i.e.,} database, data mining, artificial intelligence and computer vision.
	
	\item[$\bullet$] \textbf{CiteSeer-M10 dataset\footnote{http://citeseerx.ist.psu.edu/}} (target network) is a subset of CiteSeerX data which consists of scientific publications from 10 distinct research areas, \textit{i.e.,} agriculture, archaeology, biology, computer science, financial economics, industrial engineering, material science, petroleum chemistry, physics and social science.
\end{enumerate}

\subsubsection{Experiment Setup}\label{subsec:setupsingle}

For the evaluations, we randomly partition the dataset in the target domain into two non-overlapping sets for training and testing by nine groups of training percentages, $\{0.1,0.2,\cdots,0.9\}$. We repeat the above steps 10 times and thus obtain 10 copies of the training data and testing data. The reported experimental results are the average of the 10 runs and their variance.

The parameters of CDNR are set in line with typical values used for DeepWalk \cite{perozzi2014deepwalk}, LINE \cite{tang2015line}, Node2Vec \cite{grover2016node2vec} and Struc2Vec \cite{ribeiro2017struc2vec}. For networks in both the source domain and the target domain, let the dimensions of feature representation be $d=128$, the walk length be $l=80$, the number of walks of every source node be $k=10$, the window size be $r=10$, $workers=8$, and the search bias $\alpha$ be with $p=1$ and $q=1$. Let the learning rate $\rho$ start from 0.025 as in \cite{tang2015line} and the convergence track on 0.1 in our experiment. For Struc2Vec, let OPT1 (reducing the length of degree sequences), OPT2 (reducing the number of pairwise similarity calculations) and OPT3 (reducing the number of layers) all in values of True, and the maximum number of layers be 6. The parameters in \textit{CD2L-NodeBalance} is set as $\gamma=100$ and $\lambda=100$. In these settings, the total number of random walks over an input network is $w=SampleSize\times k$ and the size of the random walks is $w\times l$. For DeepGL \cite{rossi2018deep}, the operator is chosen from \{mean, sum, maximum, Hadamard, Weight $L^p$, RBF\} which gets best results in base feature learning; $L^p$ is set in 1; feature similarity threshold is set in 0.01; maximum depth of layer is set in 10; and convergence for feature diffusion is set in 0.01.

\subsubsection{Single-label Classification}

\begin{figure}[!htp]
	\small
	\centering
	\subfigure[Node2Vec on dblp]{\includegraphics[width=0.3\linewidth]{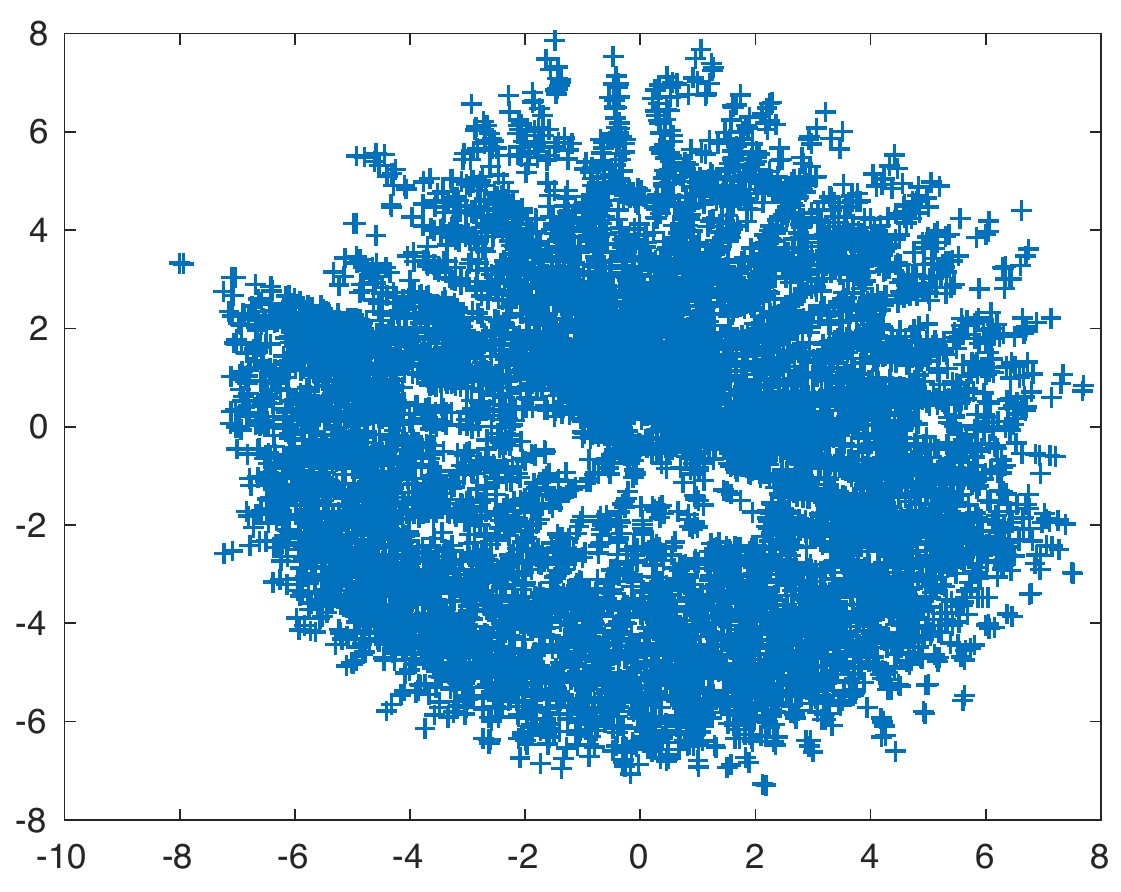}}
	\subfigure[CDNR on M10]{\includegraphics[width=0.305\linewidth]{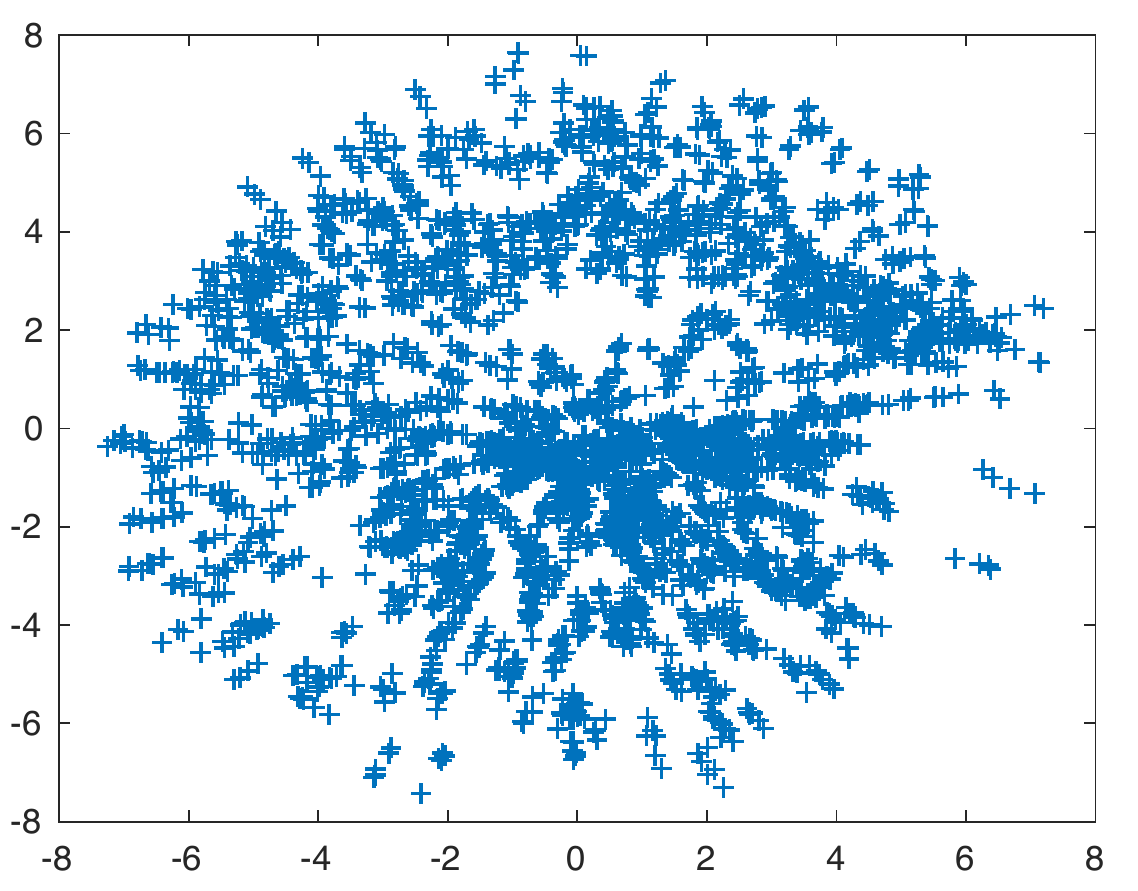}}\\
	\subfigure[PCA on M10]{\includegraphics[width=0.28\linewidth]{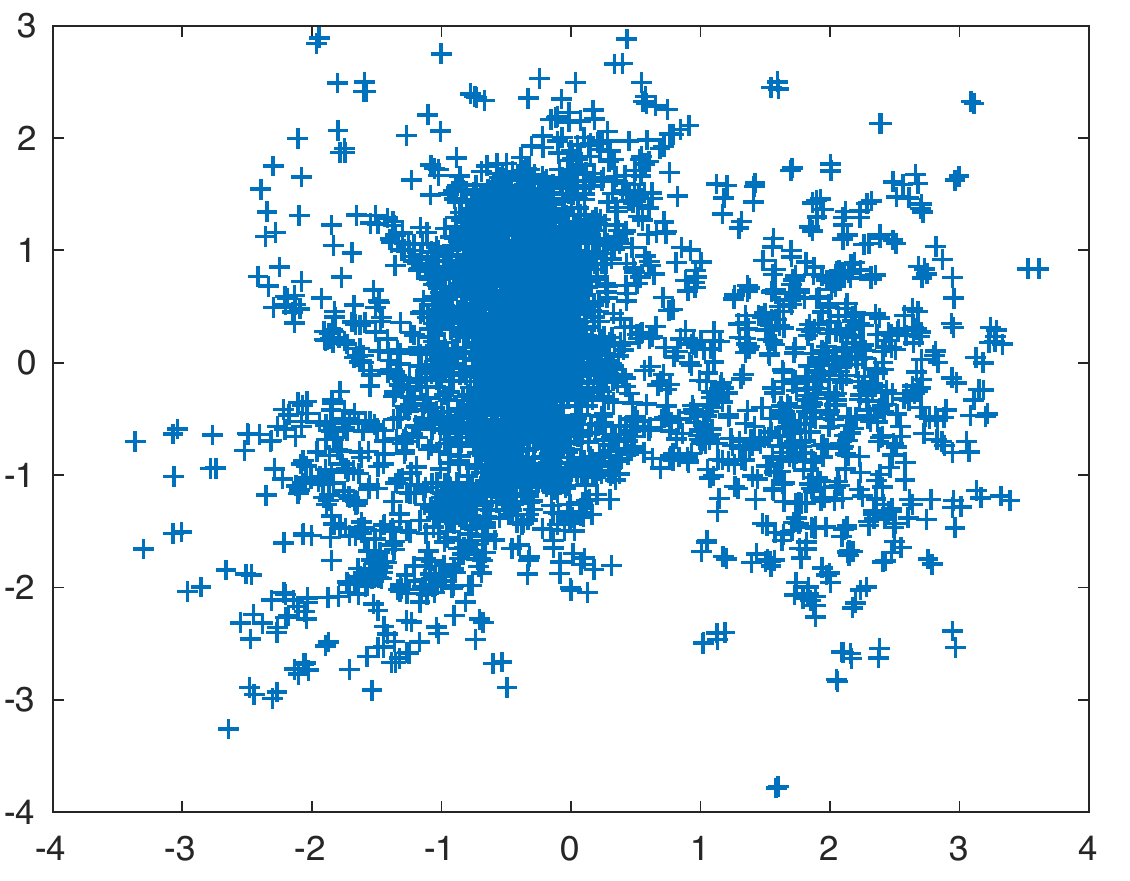}}
	\subfigure[LLE on M10]{\includegraphics[width=0.3\linewidth]{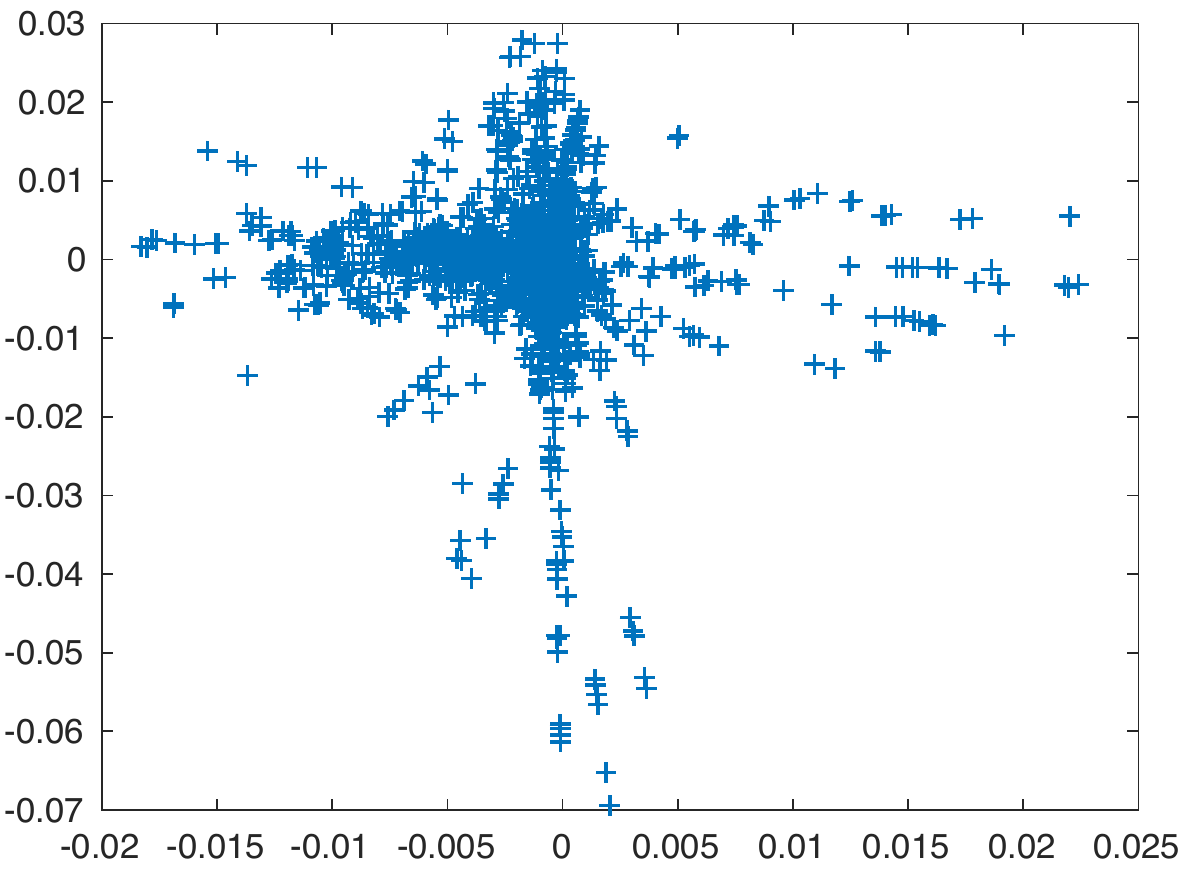}}
	\subfigure[Laplacian on M10]{\includegraphics[width=0.3\linewidth]{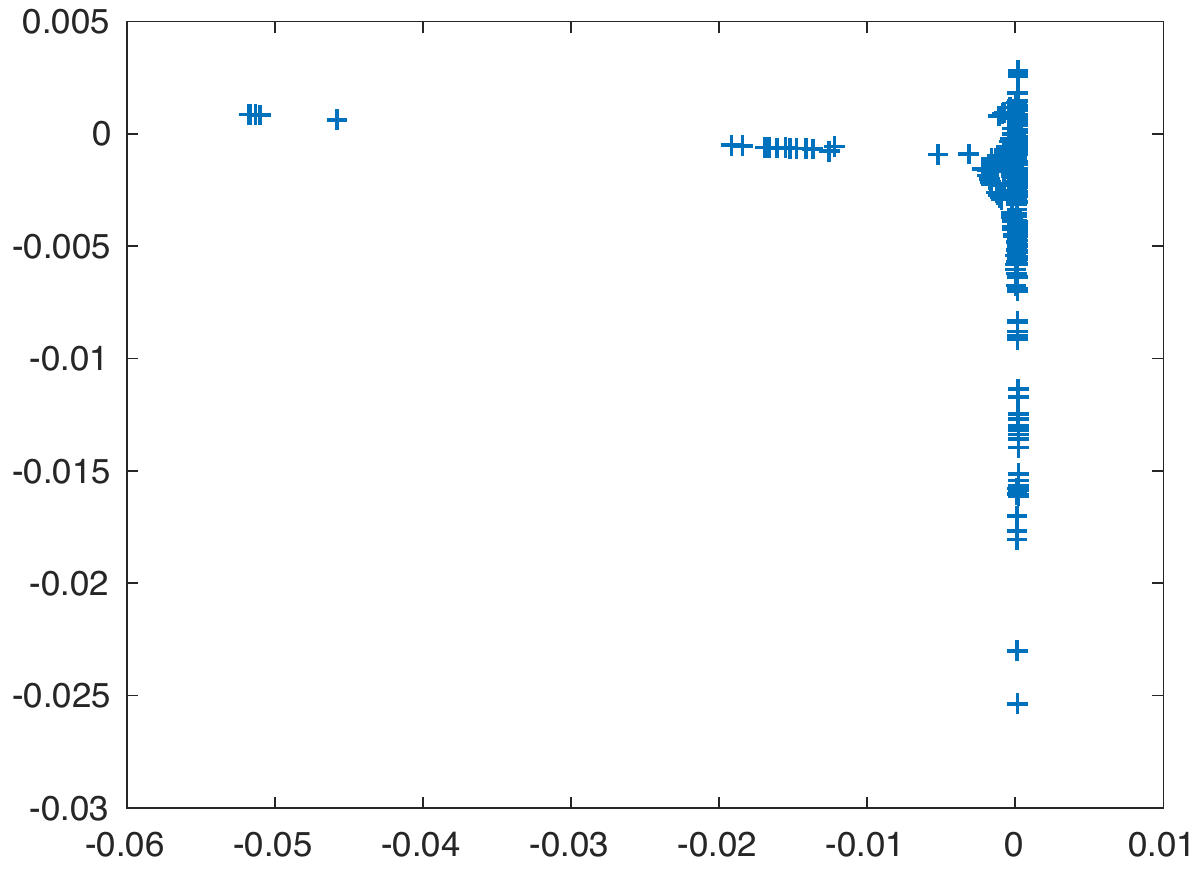}}
	\caption{Network representation on dblp and M10 in a 2-dimensional latent feature space.}
	\label{fig:d2representationsinglerw}
\end{figure}

We use Macro-F1 and Micro-F1 \cite{yang1999re} to evaluate classification performances and the results are shown in Table \ref{tab:resultsingle}. The F1 scores are designed to evaluate the effectiveness of category assignments \cite{hsu2002comparison}.

\begin{equation}\label{eq:f1}
F_1(r,p)=\frac{2rp}{r+p}
\end{equation}

We use the indicators of true positive (\textit{tp}), false positive (\textit{fp}) and false negative (\textit{fn}) to measure the standard recall ($r$) and precision ($p$). In $Micro\_F_1$, let $r=\frac{\sum{tp}}{\sum{tp}+\sum{fn}}$ and $p=\frac{\sum{tp}}{\sum{tp}+\sum{fp}}$. The Micro\_F1 score computes the global $n\times m$ binary decisions, where $n$ is the number of total test nodes, and $m$ is the number of categories of binary labels. In $Micro\_F_1$, let $r=\frac{1}{m}\sum{\frac{tp}{tp+fn}}$ and $p=\frac{1}{m}\sum{\frac{tp}{tp+fp}}$. The Macro\_F1 score computes the binary decisions on individual categories and then averages the categories.

\textbf{Representation Analysis.} Figure \ref{fig:d2representationsinglerw} (a) illustrates the feature spaces of dblp by CDNR bottom-layer random walk (Node2Vec) and Figure \ref{fig:d2representationsinglerw} (b) illustrates the feature spaces of dblp by CDNR. These two illustrations show almost the same distribution and obtain good mappings in a low dimension compared to PCA (Figure \ref{fig:d2representationsinglerw} (c)), LLE (Figure \ref{fig:d2representationsinglerw} (d)) and Laplacian (Figure \ref{fig:d2representationsinglerw} (e)) based network representations.

\begin{sidewaystable}[!htp]
	\small
	\centering
	\caption{CDNR single-label classification results on the target domain of M10.}\label{tab:resultsingle}
	\begin{tabular}{c||c|c|c|c|c|c|c|c|c|c} \hline
		& Algorithm & 10\% & 20\% & 30\% & 40\% & 50\% & 60\% & 70\% & 80\% & 90\% \\ \hline \hline
		\multirow{12}*{\rotatebox{90}{Micro-F1}} & \multirow{2}*{DeepWalk} & 0.1758 & 0.1833 & 0.1897 & 0.2049 & 0.2051 & 0.2216 & 0.2236 & 0.2420 & 0.2431 \\
		& & $\pm$0.0086 & $\pm$0.0100 & $\pm$0.0122 & $\pm$0.0126 & $\pm$0.0128 & $\pm$0.0111 & $\pm$0.0170 & $\pm$0.0133 & $\pm$0.0220 \\ \cline{2-11}
		& \multirow{2}*{LINE} & 0.2338 & 0.2362 & 0.2623 & 0.2821 & 0.3269 & 0.3244 & 0.3561 & 0.3508 & 0.4128  \\
		& & $\pm$0.0102 & $\pm$0.0170 & $\pm$0.0110 & $\pm$0.0141 & $\pm$0.0150 & $\pm$0.0087 & $\pm$0.0193 & $\pm$0.0184 & $\pm$0.0486 \\ \cline{2-11}
		& \multirow{2}*{Node2Vec} & 0.3342 & 0.4166 & 0.4714 & 0.5213 & 0.5550 & 0.5843 & 0.6216 & 0.6353 & 0.6535  \\
		& & $\pm$0.0099 & $\pm$0.0110 & $\pm$0.0153 & $\pm$0.0127 & $\pm$0.0176 & $\pm$0.0092 & $\pm$0.0215 & $\pm$0.0115 & $\pm$0.0324\\ \cline{2-11}
		& \multirow{2}*{Struc2Vec} & 0.1742 & 0.1673 & 0.1701 & 0.1622 & 0.1695 & 0.1669 & 0.1685 & 0.1626 & 0.1784 \\
		& & $\pm$0.0186 & $\pm$0.0194 & $\pm$0.0171 & $\pm$0.0215 & $\pm$0.0145 & $\pm$0.0215 & $\pm$0.0120 & $\pm$0.0188 & $\pm$0.0281 \\ \cline{2-11}
		& \multirow{2}*{DeepGL} & 0.6557 & 0.6465 & 0.6739 & 0.6600 & 0.6787 & 0.6724 & 0.6871 & 0.6786 & 0.6911 \\
		& & $\pm$0.0187 & $\pm$0.0176 & $\pm$0.0161 & $\pm$0.0197 & $\pm$0.0164 & $\pm$0.0139 & $\pm$0.0184 & $\pm$0.0160 & $\pm$0.0403 \\ \cline{2-11}		
		& \textbf{CDNR} & \textbf{0.7507} & \textbf{0.7728} & \textbf{0.8052} & \textbf{0.8245} & \textbf{0.8363} & \textbf{0.8526} & \textbf{0.8587} & \textbf{0.8772} & \textbf{0.8720} \\
		& \textbf{dblp2M10} & $\pm$\textbf{0.0143} & $\pm$\textbf{0.0114} & $\pm$\textbf{0.0154} & $\pm$\textbf{0.0074} & $\pm$\textbf{0.0051} & $\pm$\textbf{0.0116} & $\pm$\textbf{0.0128} & $\pm$\textbf{0.0173} & $\pm$\textbf{0.0179} \\ \hline
		\multirow{12}*{\rotatebox{90}{Macro-F1}} & \multirow{2}*{DeepWalk} & 0.2523 & 0.2667 & 0.2768 & 0.2945 & 0.2935 & 0.3077 & 0.3101 & 0.3294 & 0.3359 \\
		& & $\pm$0.0117 & $\pm$0.0051 & $\pm$0.0072 & $\pm$0.0120 & $\pm$0.0081 & $\pm$0.0086 & $\pm$0.0158 & $\pm$0.0123 & $\pm$0.0220\\ \cline{2-11}
		& \multirow{2}*{LINE} & 0.3160 & 0.2984 & 0.3421 & 0.3596 & 0.4070 & 0.4275 & 0.4498 & 0.4277 & 0.4773 \\
		& & $\pm$0.0113 & $\pm$0.0127 & $\pm$0.0144 & $\pm$0.0249 & $\pm$0.0382 & $\pm$0.0548 & $\pm$0.0383 & $\pm$0.0302 & $\pm$0.0486\\ \cline{2-11}
		& \multirow{2}*{Node2Vec} & 0.4326 & 0.4748 & 0.5338 & 0.5900 & 0.6092 & 0.6388 & 0.6866 & 0.6981 & 0.6568  \\
		& & $\pm$0.0147 & $\pm$0.0156 & $\pm$0.0153 & $\pm$0.0153 & $\pm$0.0290 & $\pm$0.0314 & $\pm$0.0202 & $\pm$0.0572 & $\pm$0.0261\\ \cline{2-11}
		& \multirow{2}*{Struc2Vec} & 0.2718 & 0.2129 & 0.2220 & 0.1889 & 0.2067 & 0.1871 & 0.1803 & 0.1723 & 0.1749 \\
		& & $\pm$0.0126 & $\pm$0.0092 & $\pm$0.0143 & $\pm$0.0147 & $\pm$0.0058 & $\pm$0.0150 & $\pm$0.0098 & $\pm$0.0271 & 0.0165 \\ \cline{2-11}
		& \multirow{2}*{DeepGL} & 0.6704 & 0.6003 & 0.6154 & 0.5732 & 0.5848 & 0.5607 & 0.5918 & 0.5910 & 0.5914 \\
		& & $\pm$0.0151 & $\pm$0.0222 & $\pm$0.0247 & $\pm$0.0305 & $\pm$0.0309 & $\pm$0.0264 & $\pm$0.0324 & $\pm$0.0388 & 0.0407 \\ \cline{2-11}
		& \textbf{CDNR} & \textbf{0.7558} & \textbf{0.6939} & \textbf{0.7269} & \textbf{0.7174} & \textbf{0.7301} & \textbf{0.7540} & \textbf{0.7679} & \textbf{0.7722} & \textbf{0.7745} \\
		& \textbf{dblp2M10} & $\pm$\textbf{0.0138} & $\pm$\textbf{0.0168} & $\pm$\textbf{0.0174} & $\pm$\textbf{0.0149} & $\pm$\textbf{0.0256} & $\pm$\textbf{0.0238} & $\pm$\textbf{0.0325} & $\pm$\textbf{0.0609} & $\pm$\textbf{0.0698}\\ \hline		
	\end{tabular}
\end{sidewaystable}

\textbf{Effectiveness of search priority in random walks.} In Table \ref{tab:resultsingle}, DeepWalk and Struc2Vec demonstrate worse performance than LINE, Node2Vec and our CDNR, which can be explained by their inability to reuse samples, a feat that can be easily achieved using the random walk. The outstanding performance of Node2Vec among baseline algorithms indicates that the exploration strategy is much better than the uniform random walks learned by DeepWalk and LINE. The parameter of search bias $\alpha$ adds flexibility in exploring local neighborhoods prior to the global network. The poor performance of DeepWalk and LINE mainly occurs because the network structure is rather sparse, feature noise, and contains limited information. CDNR performs best on the M10 network, as dblp is also a citation network that naturally share similar network patterns with M10. Such patterns are captured by CDNR and transfered to M10. On average, there are smaller variances in the performance of CDNR on the dblp2M10 learning task.

\textbf{Importance of information from source domain.} Table \ref{tab:resultsingle} shows that CDNR outperforms the domain-specific baseline algorithms, which use topological information from the source domain to learn the network representation in the target domain. When a top layer is working base on the \textit{CD2L-RandomWalk}, the information in the source network is transferred to the source network by adjusting the weights on the edges of the target network. This procedure achieves better performance and shows the significance of transferring topological information from the external domains.

\subsection{Experiment on Multi-label Datasets}\label{subsec:multilabel}

\subsubsection{Datasets}

\begin{table}[!htp]
	\small
	\centering
	\caption{Multi-label classification dataset statistics.}
	\label{tab:MultiLabelDataset}
	\begin{tabular}{ccccccc}
		\hline
		\multirow{2}*{\textbf{Datasets}} & \multirow{2}*{\textbf{Network}} & \textbf{Num. of} & \textbf{Num. of} & \textbf{Ave.} & \textbf{Num. of} & \multirow{2}*{\textbf{Labels}} \\
		& & \textbf{Nodes} & \textbf{Edges} & \textbf{Degree} & \textbf{Categories} &  \\
		\hline
		Blog3 & Social & 10,312 & 333,983 & 64.776 & 39 & Interests \\
		Facebook & Social & 4,039 & 88,234 & 43.691 & 10 & Groups \\
		PPI & Biological & 3,890 & 37,845 & 19.609 & 50 & States \\
		arXivCit-HepPh & Citation & 34,546 & 421,578 & 24.407 & 11 & Years \\
		arXivCit-HepTh & Citation & 27,777 & 352,807 & 25.409 & 11 & Years \\
		\hline
	\end{tabular}
\end{table}

\begin{figure}[!htp]
	\small
	\centering
	\subfigure[\tiny Blog3]{\includegraphics[width=0.16\linewidth]{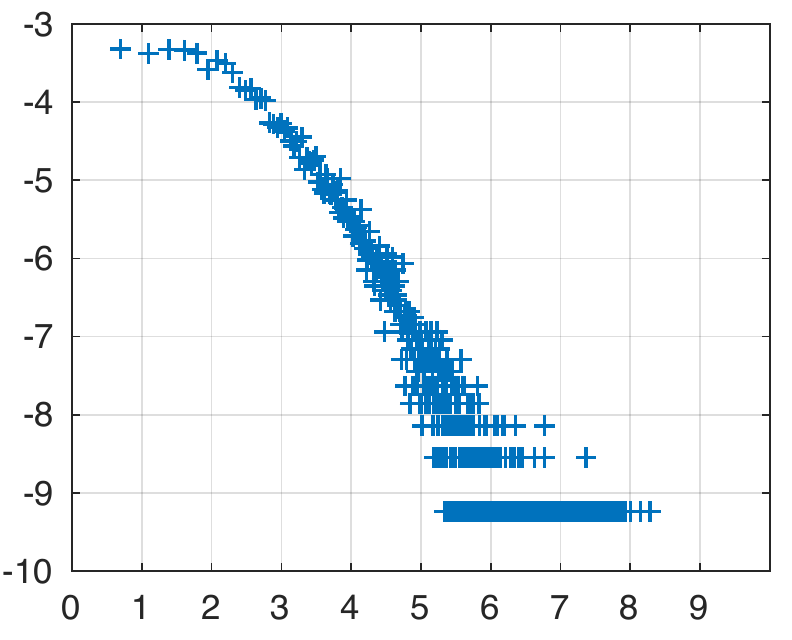}}
	\subfigure[\tiny Blog3 RW]{\includegraphics[width=0.16\linewidth]{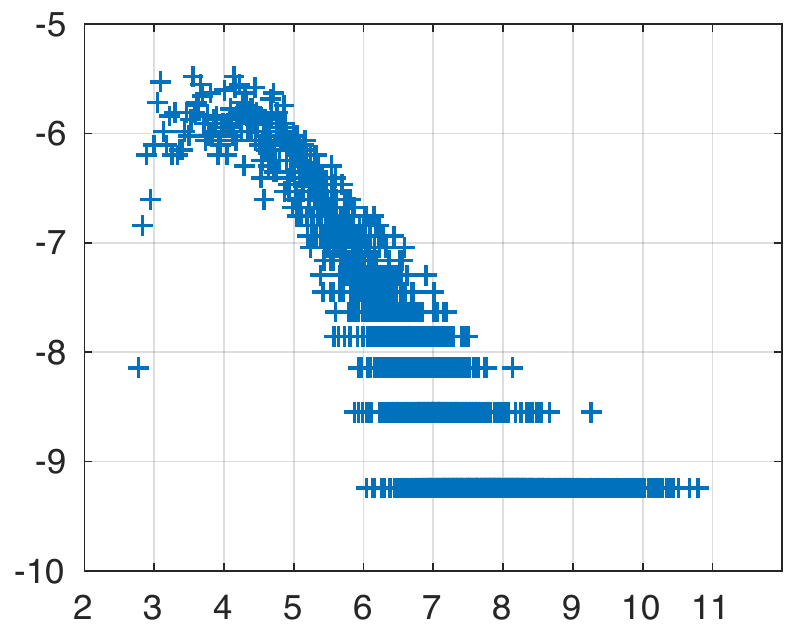}}
	\subfigure[\tiny Facebook]{\includegraphics[width=0.16\linewidth]{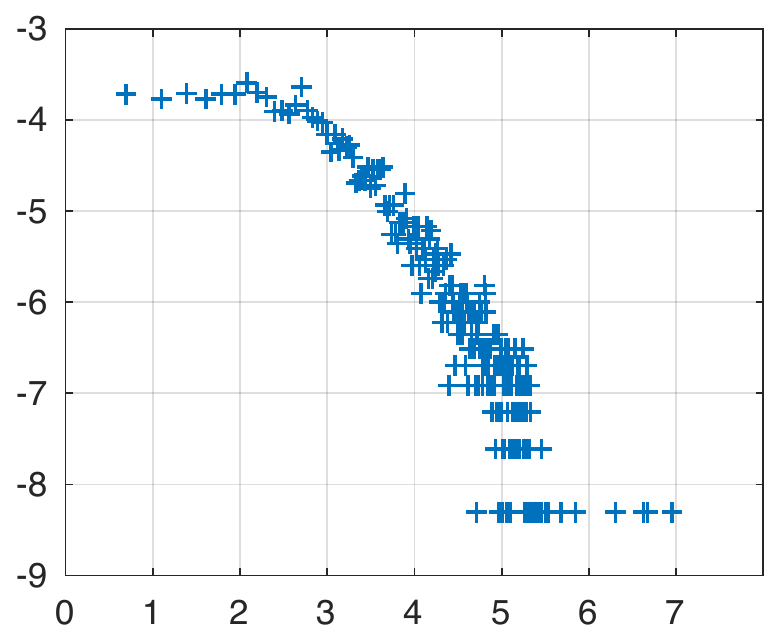}}
	\subfigure[\tiny Facebook RW]{\includegraphics[width=0.16\linewidth]{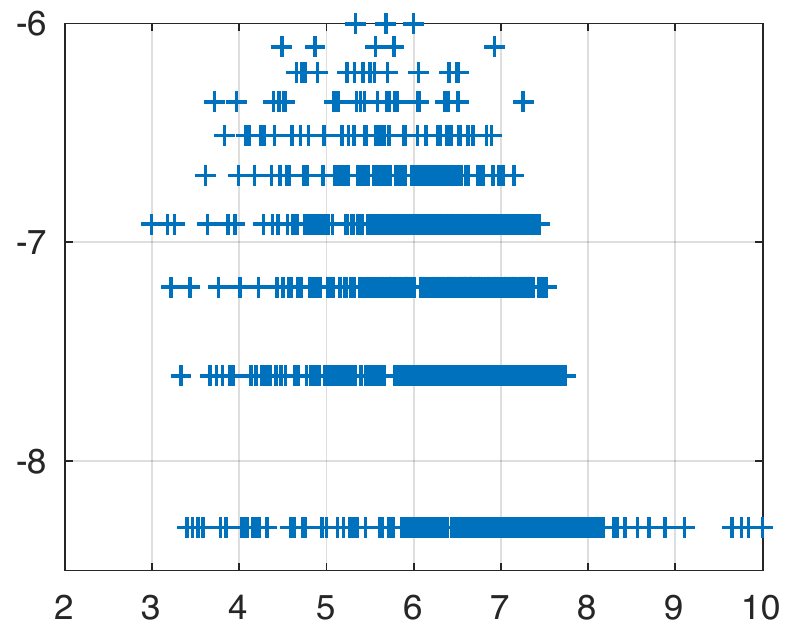}}
	\subfigure[\tiny PPI]{\includegraphics[width=0.16\linewidth]{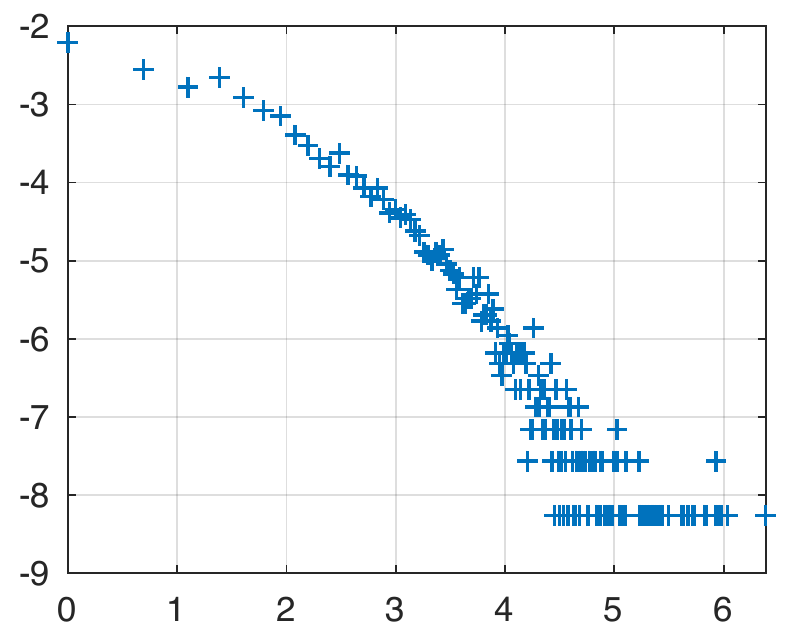}} \\
	\subfigure[\tiny arXivHepPh]{\includegraphics[width=0.16\linewidth]{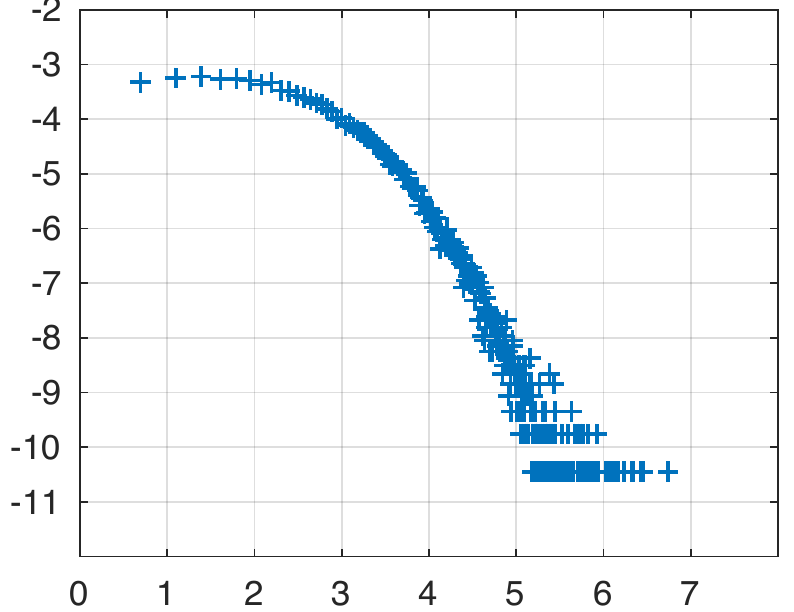}}
	\subfigure[\tiny arXivHepPh RW]{\includegraphics[width=0.16\linewidth]{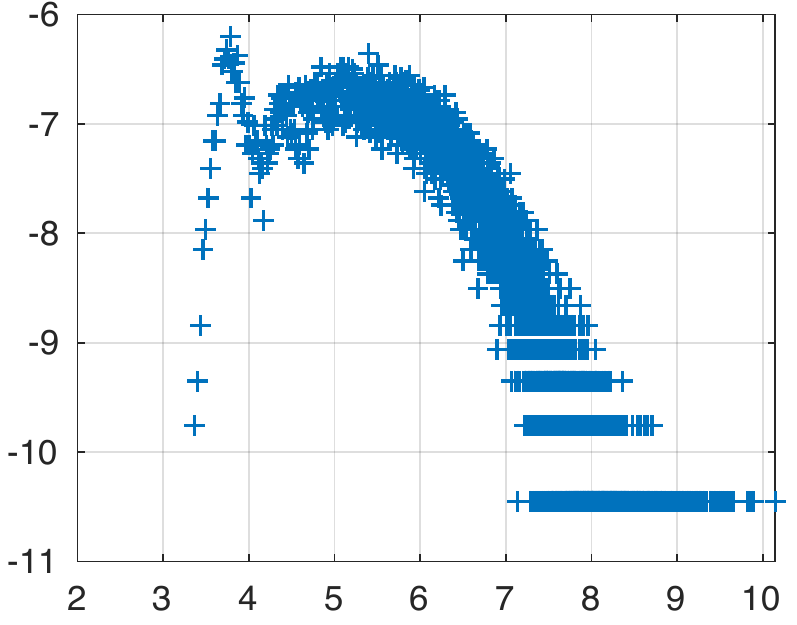}}
	\subfigure[\tiny arXivHepTh]{\includegraphics[width=0.16\linewidth]{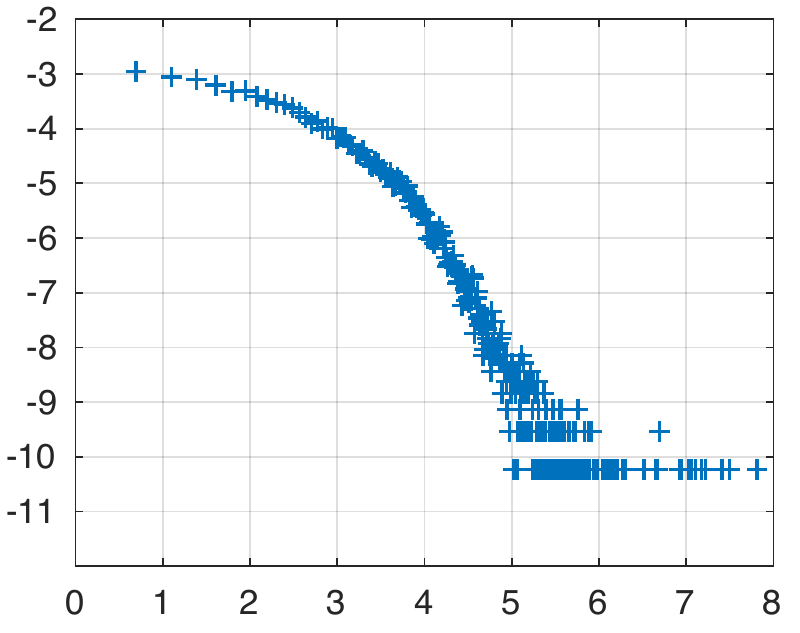}}
	\subfigure[\tiny arXivHepTh RW]{\includegraphics[width=0.16\linewidth]{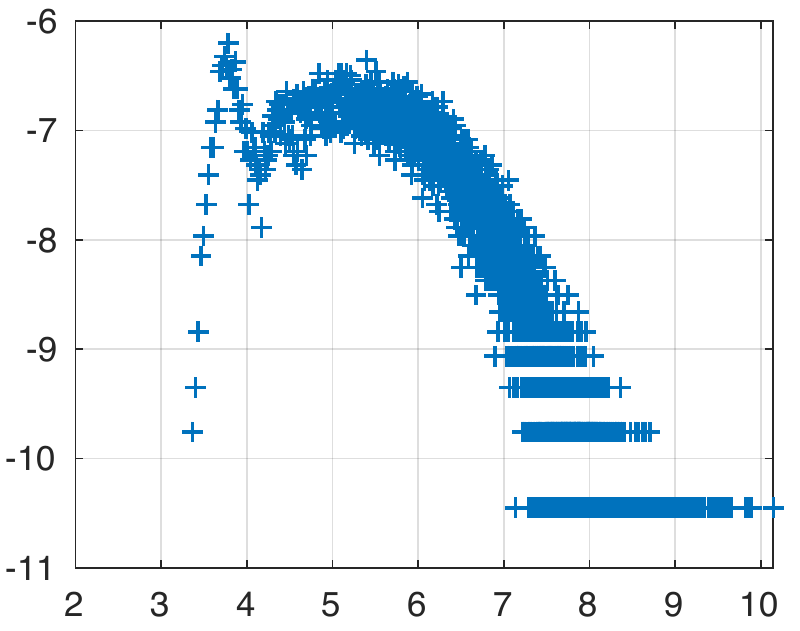}}
	\subfigure[\tiny PPI RW]{\includegraphics[width=0.16\linewidth]{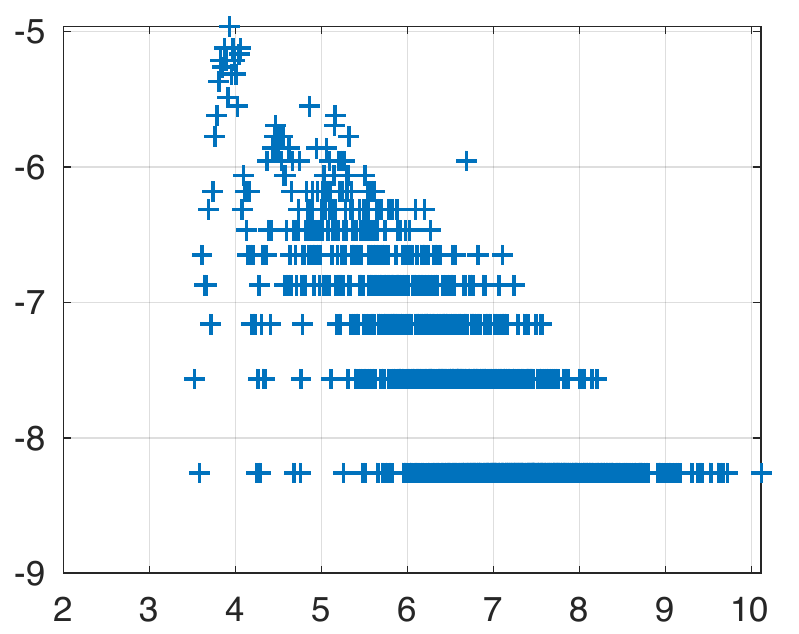}}
	\caption{Power-law distributions of multi-label classification datasets and their random walks.}
	\label{fig:powerlawmulti}
\end{figure}

We select five real-world large-scale networks of different kinds as the experimental datasets, consisting of three online social networks (Blog3, Facebook), two citation networks (arXivCit-HepPh, arXivCit-HepTh) and one biological network (PPI). All of them are for the multi-class multi-label classification problem. In the online social networks, nodes represent users and the users' relationships are denoted as edges. In the citation networks, papers are denoted as nodes and edges describe the citations in this experiment. In the biological network, genes are denoted as nodes and edges represent the relationships between the genes.

\begin{enumerate}
	
	\item[$\bullet$] \textbf{Blog3 (BlogCatalog3) dataset\footnote{http://socialcomputing.asu.edu/datasets/BlogCatalog3}} is a social blog directory which manages bloggers and their blogs. Both the contact network and selected group membership information is included. The network has 10,312 nodes, 333,983 undirected edges and 39 different labels. Nodes are classified according to the interests of bloggers.
	
	\item [$\bullet$] \textbf{Facebook dataset\footnote{https://snap.stanford.edu/data/egonets-Facebook.html}} consists of circles (i.e., friends lists) from Facebook. This dataset contains user profiles as node features, and circles as edge features and ego networks. The network has 4,039 nodes, 88,234 undirected edges and 10 different labels representing groups of users.
	
	\item[$\bullet$] \textbf{PPI (Protein-Protein Interactions) dataset\footnote{https://downloads.thebiogrid.org/BioGRID}} is a subgraph of the PPI network for Homo Sapiens, which obtains labels from hallmark gene sets and represents biological states. The network has 3,890 nodes, 76,584 undirected edges and 50 different labels.
	
	\item[$\bullet$] \textbf{arXivCit-HepPh (arXiv High-energy Physics Citation Network) dataset\footnote{http://snap.stanford.edu/data/cit-HepPh.html} and arXivCit-HepTh (arXiv High-energy Physics Theory Citation Network) dataset\footnote{http://snap.stanford.edu/data/cit-HepTh.html}} are abstracted from the e-print arXiv. arXivCit-HepPh covers all the citations within a dataset of 34,546 papers (regarded as nodes) with 421,578 directed edges. arXivCit-HepTh covers all the citations within a dataset of 27,777 papers (regarded as nodes) with 352,807 directed edges. If a paper $v_i$ cites paper $v_j$, the graph contains a directed edge from $v_i$ to $v_j$. The data consist of papers from the period January 1993 to April 2003, categorized by year.
\end{enumerate}

The networks chosen in the experiment follow the power-law distribution \cite{adamic2000power}, as do the random walks on the networks \cite{perozzi2014deepwalk}, as shown in Figure \ref{fig:powerlawmulti}.

\subsubsection{Experiment Setup}

\begin{table}[!htp]
	\small
	\centering
	\addtolength{\tabcolsep}{50pt}
	\caption{Networks selected as the source domain and target domain for CDNR by distance.}
	\label{Table:domians}
	\begin{tabular}{c|c}
		\hline
		\textbf{Source Domain} &  \textbf{Target Domain} \\
		\hline
		Blog3 & PPI \\
		arXivCit-HepTh & PPI \\
		arXivCit-HepPh & PPI \\
		Facebook & PPI \\ \hline
		Blog3 & Facebook \\
		\hline
	\end{tabular}
\end{table}

This experiment summarizes the network statistics in Table \ref{tab:MultiLabelDataset}. Node degree reflects the connection capability of the node. A network is selected as a source domain or a target domain follows $|V^s|>|V^t|$ and $\langle deg^s \rangle>\langle deg^t \rangle$. These selections are shown in Table \ref{Table:domians}. The experiment setup for the multi-label classification evaluation is as same as the setup in the single-label dataset experiment.

\subsubsection{Multi-label Classification}

In the multi-label classification setting, every node is assigned one or more labels from a finite set $Y$. In the training phase of the CDNR node feature representations, we observe a fraction of the nodes and all their labels, and predict the labels for the remaining nodes. The multi-label classification in our experiment inputs the network representations to a \textit{one-against-all} linear SVM classifier \cite{hsu2002comparison}. We use the F1 score of Macro-F1 and Micro-F1 to compare performance \cite{yang1999re} in Tables \ref{tab:CDRNPPImicro}-\ref{tab:CDNRfacebookmacro}.

\begin{sidewaystable}[!htp]
	\centering
	\small
	\caption{CDRN multi-label classification results of Micro-F1 on the target domain network of PPI.}\label{tab:CDRNPPImicro}
	\begin{tabular}{c|c|c|c|c|c|c|c|c|c} \hline
		Algorithm & 10\% & 20\% & 30\% & 40\% & 50\% & 60\% & 70\% & 80\% & 90\% \\ \hline \hline
		\multirow{2}*{DeepWalk} & 0.2849 & 0.2854 & 0.2845 & 0.2803 & 0.2725 & 0.2736 & 0.2629 & 0.2778 & 0.2621  \\
		& $\pm$0.0181 & $\pm$0.0116 & $\pm$0.0193 & $\pm$0.0170 & $\pm$0.0168 & $\pm$0.0200 & $\pm$0.0241 & $\pm$0.0215 & $\pm$0.0344 \\ \hline
		\multirow{2}*{LINE} & 0.2900 & 0.2772 & 0.2807 & 0.2715 & 0.2702 & 0.2649 & 0.2710 & 0.2494 & 0.2398  \\
		& $\pm$0.0062 & $\pm$0.0077 & $\pm$0.0083 & $\pm$0.0104 & $\pm$0.0113 & $\pm$0.0166 & $\pm$0.0163 & $\pm$0.0251 & $\pm$0.0195 \\ \hline
		\multirow{2}*{Node2Vec} & 0.3073 & 0.2955 & 0.3024 & 0.3028 & 0.3028 & 0.2995 & 0.3021 & 0.2967 & 0.3005  \\
		& $\pm$0.0171 & $\pm$0.0104 & $\pm$0.0139 & $\pm$0.0120 & $\pm$0.0102 & $\pm$0.0186 & $\pm$0.0288 & $\pm$0.0197 & $\pm$0.0283 \\ \hline
		\multirow{2}*{Struc2Vec} & 0.2693 & 0.2713 & 0.2696 & 0.2515 & 0.2603 & 0.2499 & 0.2493 & 0.2419 & 0.2338 \\
		& $\pm$0.0228 & $\pm$0.0187 & $\pm$0.0188 & $\pm$0.0187 & $\pm$0.0133 & $\pm$0.0212 & $\pm$0.0148 & $\pm$0.0156 & $\pm$0.0287 \\ \hline
		\multirow{2}*{DeepGL} & 0.3055 & 0.3063 & 0.3028 & 0.2947 & 0.2987 & 0.2975 & 0.2911 & 0.2890 & 0.2764 \\
		& $\pm$0.0062 & $\pm$0.0083 & $\pm$0.0083 & $\pm$0.0054 & $\pm$0.0063 & $\pm$0.0128 & $\pm$0.0128 & $\pm$0.0180 & $\pm$0.0178 \\ \hline
		\textbf{CDNR} & \multirow{2}*{0.3386} & \multirow{2}*{0.3390} & \multirow{2}*{0.3423} & \multirow{2}*{0.3420} & \multirow{2}*{0.3404} & \multirow{2}*{0.3414} & \multirow{2}*{0.3350} & \multirow{2}*{0.3371} & \multirow{2}*{0.3312} \\
		\textbf{Blog3} & \multirow{2}*{$\pm$0.0062} & \multirow{2}*{$\pm$0.0086} & \multirow{2}*{$\pm$0.0061} & \multirow{2}*{$\pm$0.0072} & \multirow{2}*{$\pm$0.0079} & \multirow{2}*{$\pm$0.0073} & \multirow{2}*{$\pm$0.0125} & \multirow{2}*{$\pm$0.0199} & \multirow{2}*{$\pm$0.0238} \\
		\textbf{2PPI}  & & & & & & & & & \\ \hline
		\textbf{CDNR} &  & & & & & & & &  \\
		\textbf{arXivCit} & 0.3412 & 0.3425 & 0.3410 & 0.3431 & 0.3460 & 0.3474 & 0.3431 & 0.3429 & 0.3330 \\
		\textbf{-HepPh}  & $\pm$0.0041 & $\pm$0.0075 & $\pm$0.0052 & $\pm$0.0057 & $\pm$0.0069 & $\pm$0.0112 & $\pm$0.0103 & $\pm$0.0107 & $\pm$0.0192 \\
		\textbf{2PPI}  & & & & & & & & & \\ \hline
		\textbf{CDNR} & & & & & & & & &  \\
		\textbf{arXivCit} & \textbf{0.3420} & 0.3426 & 0.3434 & \textbf{0.3462} & 0.3441 & \textbf{0.3553} & \textbf{0.3450} & \textbf{0.3457} & \textbf{0.3521} \\
		\textbf{-HepTh} & \textbf{$\pm$0.0036} & $\pm$0.0057 & $\pm$0.0044 & \textbf{$\pm$0.0049} & $\pm$0.0042 & \textbf{$\pm$0.0101} & \textbf{$\pm$0.0106} & \textbf{$\pm$0.0150} & \textbf{$\pm$0.0260} \\
		\textbf{2PPI}  & & & & & & & & & \\ \hline
		\textbf{CDNR} & \multirow{2}*{0.3415} & \multirow{2}*{\textbf{0.3442}} & \multirow{2}*{\textbf{0.3454}} & \multirow{2}*{0.3448} & \multirow{2}*{\textbf{0.3468}} & \multirow{2}*{0.3410} & \multirow{2}*{0.3447} & \multirow{2}*{0.3443} & \multirow{2}*{0.3444} \\
		\textbf{Facebook} & \multirow{2}*{$\pm$0.0053} & \multirow{2}*{\textbf{$\pm$0.0035}} & \multirow{2}*{\textbf{$\pm$0.0035}} & \multirow{2}*{$\pm$0.0066} & \multirow{2}*{\textbf{$\pm$0.0065}} & \multirow{2}*{$\pm$0.0102} & \multirow{2}*{$\pm$0.0119} & \multirow{2}*{$\pm$0.0151} & \multirow{2}*{$\pm$0.0189} \\
		\textbf{2PPI}  & & & & & & & & & \\ \hline
	\end{tabular}
\end{sidewaystable}

\begin{sidewaystable}[!htp]
	\centering
	\small
	\caption{CDRN multi-label classification results of Macro-F1 on the target domain network of PPI.}\label{tab:CDRNPPImacro}
	\begin{tabular}{c|c|c|c|c|c|c|c|c|c} \hline
		Algorithm & 10\% & 20\% & 30\% & 40\% & 50\% & 60\% & 70\% & 80\% & 90\% \\ \hline \hline
		\multirow{2}*{DeepWalk} & 0.3416 & 0.3378 & 0.3364 & 0.3406 & 0.3306 & 0.3336 & 0.2949 & 0.2825 & 0.2041 \\
		& $\pm$0.0140 & $\pm$0.0138 & $\pm$0.0208 & $\pm$0.0171 & $\pm$0.0159 & $\pm$0.0241 & $\pm$0.0288 & $\pm$0.0185 & $\pm$0.0386 \\ \hline
		\multirow{2}*{LINE} & 0.3058 & 0.3003 & 0.3008 & 0.2940 & 0.2868 & 0.2826 & 0.2733 & 0.2462 & 0.1822  \\
		& $\pm$0.0094 & $\pm$0.0113 & $\pm$0.0069 & $\pm$0.0120 & $\pm$0.0138 & $\pm$0.0176 & $\pm$0.0173 & $\pm$0.0262 & $\pm$0.0198 \\ \hline
		\multirow{2}*{Node2Vec} & 0.3490 & 0.3442 & 0.3510 & 0.3500 & 0.3432 & 0.3414 & 0.3274 & 0.3006 & 0.2310 \\
		& $\pm$0.0193 & $\pm$0.0141 & $\pm$0.0205 & $\pm$0.0126 & $\pm$0.0140 & $\pm$0.0201 & $\pm$0.0240 & $\pm$0.0248 & $\pm$0.0385 \\ \hline
		\multirow{2}*{Struc2Vec} & 0.2892 & 0.2926 & 0.3019 & 0.2784 & 0.2851 & 0.2626 & 0.2589 & 0.2399 & 0.1712 \\
		& $\pm$0.0197 & $\pm$0.0232 & $\pm$0.0227 & $\pm$0.0267 & $\pm$0.0152 & $\pm$0.0202 & $\pm$0.0177 & $\pm$0.0287 & $\pm$0.0262 \\ \hline
		\multirow{2}*{DeepGL} & 0.3213 & 0.3290 & 0.3235 & 0.3155 & 0.3136 & 0.3086 & 0.2970 & 0.2783 & 0.2115 \\
		& $\pm$0.0065 & $\pm$0.0072 & $\pm$0.0107 & $\pm$0.0067 & $\pm$0.0095 & $\pm$0.0117 & $\pm$0.0147 & $\pm$0.0220 & $\pm$0.0132 \\ \hline
		\textbf{CDNR} & \multirow{2}*{0.3519} & \multirow{2}*{0.3551} & \multirow{2}*{0.3539} & \multirow{2}*{0.3514} & \multirow{2}*{0.3469} & \multirow{2}*{0.3431} & \multirow{2}*{0.3282} & \multirow{2}*{0.3063} & \multirow{2}*{0.2389} \\
		\textbf{Blog3} & \multirow{2}*{$\pm$0.0108} & \multirow{2}*{$\pm$0.0105} & \multirow{2}*{$\pm$0.0073} & \multirow{2}*{$\pm$0.0060} & \multirow{2}*{$\pm$0.0097} & \multirow{2}*{$\pm$0.0060} & \multirow{2}*{$\pm$0.0154} & \multirow{2}*{$\pm$0.0223} & \multirow{2}*{$\pm$0.0276} \\
		\textbf{2PPI}  & & & & & & & & & \\ \hline
		\textbf{CDNR} & & & & & & & & &  \\
		\textbf{arXivCit} & 0.3532 & 0.3582 & 0.3536 & 0.3509 & 0.3531 & 0.3456 & 0.3360 & 0.3130 & 0.2512 \\
		\textbf{-HepPh} & $\pm$0.0106 & $\pm$0.0099 & $\pm$0.0085 & $\pm$0.0057 & $\pm$0.0098 & $\pm$0.0085 & $\pm$0.0128 & $\pm$0.0144 & $\pm$0.0270 \\
		\textbf{2PPI}  & & & & & & & & & \\ \hline
		\textbf{CDNR} & & & & & & & & &  \\
		\textbf{arXivCit} & 0.3570 & 0.3575 & 0.3568 & \textbf{0.3565} & 0.3523 & \textbf{0.3556} & 0.3368 & \textbf{0.3234} & \textbf{0.2682} \\
		\textbf{-HepTh} & $\pm$0.0079 & $\pm$0.0091 & $\pm$0.0044 & \textbf{$\pm$0.0091} & $\pm$0.0090 & \textbf{$\pm$0.0137} & $\pm$0.0150 & \textbf{$\pm$0.0249} & \textbf{$\pm$0.0330}\\
		\textbf{2PPI}  & & & & & & & & & \\ \hline
		\textbf{CDNR} & \multirow{2}*{\textbf{0.3576}} & \multirow{2}*{\textbf{0.3595}} & \multirow{2}*{\textbf{0.3574}} & \multirow{2}*{0.3553} & \multirow{2}*{\textbf{0.3573}} & \multirow{2}*{0.3432} & \multirow{2}*{\textbf{0.3423}} & \multirow{2}*{0.3164} & \multirow{2}*{0.2578} \\
		\textbf{Facebook} & \multirow{2}*{\textbf{$\pm$0.0086}} & \multirow{2}*{\textbf{$\pm$0.0065}} & \multirow{2}*{\textbf{$\pm$0.0052}} & \multirow{2}*{$\pm$0.0068} & \multirow{2}*{\textbf{$\pm$0.0080}} & \multirow{2}*{$\pm$0.0079} & \multirow{2}*{\textbf{$\pm$0.0145}} & \multirow{2}*{$\pm$0.0164} & \multirow{2}*{$\pm$0.0217} \\
		\textbf{2PPI}  & & & & & & & & & \\ \hline
	\end{tabular}
\end{sidewaystable}

\begin{sidewaystable}[!htp]
	\small
	\centering \caption{CDNR multi-label classification results of Micro-F1 on the target domain network of Facebook.}\label{tab:CDNRfacebookmicro}
	\begin{tabular}{c|c|c|c|c|c|c|c|c|c} \hline
		Algorithm & 10\% & 20\% & 30\% & 40\% & 50\% & 60\% & 70\% & 80\% & 90\% \\ \hline \hline
		\multirow{2}*{DeepWalk} & 0.8078 & 0.8727 & 0.8933 & 0.9050 & 0.9153 & 0.9198 & 0.9307 & 0.9301 & 0.9334 \\
		& $\pm$0.0449 & $\pm$0.0177 & $\pm$0.0062 & $\pm$0.0059 & $\pm$0.0060 & $\pm$0.0061 & $\pm$0.0039 & $\pm$0.0103 & $\pm$0.0175 \\ \hline
		\multirow{2}*{LINE} & 0.4627 & 0.4654 & 0.4719 & 0.4739 & 0.4765 & 0.4761 & 0.4760 & 0.4787 & 0.4755  \\
		& $\pm$0.0026 & $\pm$0.0104 & $\pm$0.0026 & $\pm$0.0035 & $\pm$0.0035 & $\pm$0.0033 & $\pm$0.0067 & $\pm$0.0066 & $\pm$0.0075 \\ \hline
		\multirow{2}*{Node2Vec} & 0.9352 & 0.9401 & 0.9398 & 0.9419 & 0.9442 & 0.9454 & 0.9468 & 0.9466 & 0.9502  \\
		& $\pm$0.0072 & $\pm$0.0032 & $\pm$0.0051 & $\pm$0.0047 & $\pm$0.0057 & $\pm$0.0063 & $\pm$0.0092 & $\pm$0.0079 & $\pm$0.0098 \\ \hline
		\multirow{2}*{Struc2Vec} & 0.4152 & 0.4521 & 0.4716 & 0.4994 & 0.5161 & 0.5381 & 0.5461 & 0.5639 & 0.5530 \\
		& $\pm$0.0237 & $\pm$0.0144 & $\pm$0.0061 & $\pm$0.0059 & $\pm$0.0078 & $\pm$0.0096 & $\pm$0.0115 & $\pm$0.0241 & $\pm$0.0175 \\ \hline
		\multirow{2}*{DeepGL} & 0.9535 & 0.9483 & 0.9531 & 0.9515 & 0.9552 & 0.9546 & 0.9555 & 0.9612 & \textbf{0.9640} \\
		& $\pm$0.0161 & $\pm$0.0114 & $\pm$0.0059 & $\pm$0.0099 & $\pm$0.0087 & $\pm$0.0087 & $\pm$0.0039 & $\pm$0.0089 & \textbf{$\pm$0.0072} \\ \hline
		\textbf{CDNR} & \multirow{2}*{\textbf{0.9584}} & \multirow{2}*{\textbf{0.9550}} & \multirow{2}*{\textbf{0.9561}} & \multirow{2}*{\textbf{0.9565}} & \multirow{2}*{\textbf{0.9568}} & \multirow{2}*{\textbf{0.9617}} & \multirow{2}*{\textbf{0.9606}} & \multirow{2}*{\textbf{0.9627}} & \multirow{2}*{0.9623} \\
		\textbf{Blog3} & \multirow{2}*{\textbf{$\pm$0.0038}} & \multirow{2}*{\textbf{$\pm$0.0036}} & \multirow{2}*{\textbf{$\pm$0.0049}} & \multirow{2}*{\textbf{$\pm$0.0044}} & \multirow{2}*{\textbf{$\pm$0.0032}} & \multirow{2}*{\textbf{$\pm$0.0050}} & \multirow{2}*{\textbf{$\pm$0.0042}} & \multirow{2}*{\textbf{$\pm$0.0067}} & \multirow{2}*{$\pm$0.0085} \\
		\textbf{2Facebook}  & & & & & & & & & \\ \hline
	\end{tabular}
	\centering \caption{CDNR multi-label classification results of Macro-F1 on the target domain network of Facebook.}\label{tab:CDNRfacebookmacro}
	\begin{tabular}{c|c|c|c|c|c|c|c|c|c} \hline
		Algorithm & 10\% & 20\% & 30\% & 40\% & 50\% & 60\% & 70\% & 80\% & 90\% \\ \hline \hline
		\multirow{2}*{DeepWalk} & 0.7655 & 0.7915 & 0.7858 & 0.8052 & 0.7902 & 0.8138 & 0.8213 & 0.7678 & 0.7822 \\
		& $\pm$0.0185 & $\pm$0.0242 & $\pm$0.0331 & $\pm$0.0308 & $\pm$0.0306 & $\pm$0.0327 & $\pm$0.0504 & $\pm$0.0317 & $\pm$0.0378 \\ \hline
		\multirow{2}*{LINE} & 0.5063 & 0.5040 & 0.5083 & 0.5129 & 0.5091 & 0.5040 & 0.5020 & 0.4981 & 0.4961  \\
		& $\pm$0.0053 & $\pm$0.0189 & $\pm$0.0093 & $\pm$0.0061 & $\pm$0.0092 & $\pm$0.0077 & $\pm$0.0137 & $\pm$0.0117 & $\pm$0.0109 \\ \hline
		\multirow{2}*{Node2Vec} & 0.8310 & 0.8331 & 0.8206 & \textbf{0.8373} & 0.8343 & 0.8214 & 0.8192 & 0.8018 & 0.8104 \\
		& $\pm$0.0256 & $\pm$0.0226 & $\pm$0.0262 & \textbf{$\pm$0.0359} & $\pm$0.0354 & $\pm$0.0479 & $\pm$0.0487 & $\pm$0.0277 & $\pm$0.0498 \\ \hline
		\multirow{2}*{Struc2Vec} & 0.3701 & 0.3937 & 0.3926 & 0.4160 & 0.4377 & 0.4525 & 0.4532 & 0.4755 & 0.4583 \\
		& $\pm$0.0156 & $\pm$0.0157 & $\pm$0.0174 & $\pm$0.0155 & $\pm$0.0235 & $\pm$0.0131 & $\pm$0.0144 & $\pm$0.0260 & $\pm$0.0347 \\ \hline
		\multirow{2}*{DeepGL} & \textbf{0.8810} & 0.8660 & 0.8724 & 0.8748 & 0.8794 & 0.8856 & 0.8578 & \textbf{0.8732} & \textbf{0.8757} \\
		& $\pm$\textbf{0.0330} & $\pm$0.0328 & $\pm$0.0381 & $\pm$0.0343 & $\pm$0.0395 & $\pm$0.0313 & $\pm$0.0350 & \textbf{$\pm$0.0514} & \textbf{$\pm$0.0537} \\ \hline
		\textbf{CDNR} & \multirow{2}*{0.8749} & \multirow{2}*{\textbf{0.8831}} & \multirow{2}*{\textbf{0.8866}} & \multirow{2}*{\textbf{0.8766}} & \multirow{2}*{\textbf{0.8890}} & \multirow{2}*{\textbf{0.8876}} & \multirow{2}*{\textbf{0.8910}} & \multirow{2}*{0.8443} & \multirow{2}*{0.8415} \\
		\textbf{Blog3} & \multirow{2}*{$\pm$0.0301} & \multirow{2}*{\textbf{$\pm$0.0135}} & \multirow{2}*{\textbf{$\pm$0.0304}} & \multirow{2}*{\textbf{$\pm$0.0360}} & \multirow{2}*{\textbf{$\pm$0.0291}} & \multirow{2}*{\textbf{$\pm$0.0294}} & \multirow{2}*{\textbf{$\pm$0.0334}} & \multirow{2}*{$\pm$0.0468} & \multirow{2}*{$\pm$0.0482} \\
		\textbf{2Facebook}  & & & & & & & & & \\ \hline	
	\end{tabular}
\end{sidewaystable}

\textbf{Experimental results from the algorithmic perspective.} A general observation drawn from the results is that the learned feature representations from other networks improve or maintain performance compared to the domain-specific network representation baseline algorithms. CDNR outperforms DeepWalk, LINE, Node2Vec, Struc2Vec and DeepGL in all datasets with a gain of 19.29\%, 49.57\%, 15.66\%, 58.83\% and 10.06\%, respectively. CDNR outperforms DeepWalk, LINE, Node2Vec and Struc2Vec on the PPI dataset and the Facebook dataset in 100\% of the experiment, and outperforms DeepGL on the PPI dataset in 100\% and the Facebook dataset in 88.89\% of the experiment. The losses of CDNR to DeepGL on the training percentages of \{80\%,90\%\} might caused by classifier and training sample selection and NN-based DeepGL shows robustness than other algorithms. 

\textbf{Experimental results from the dataset perspective.} The general results on the PPI dataset (Tables \ref{tab:CDRNPPImicro} and \ref{tab:CDRNPPImacro}) reflect the difficulty of cross-domain learning. Considering the domain similarities, a cross-domain adaption from either the social networks or the citation networks to the biological network as shown in our experiment would not be recommended in transfer learning. However, CDNR is capable of capturing useful structural information from network topologies and removing noise from the source domain networks in an unsupervised feature-learning environment, so CDNR on PPI still shows a slight improvement and almost retains its representation performances. Therefore, cross-domain network knowledge transfer learning works in unsupervised network representations. CDNR is less influenced by domain selections when the transferable knowledge is mainly contributed by network topologies.

Examining the results in detail shows that the source domain networks of arXivCit-HepTh and Facebook provide a larger volume of information to the PPI target domain network than other pairs of CDNR experiments, which promote knowledge transfer across domains. The citation networks of arXivCit-HepPh and arXivCit-HepTh transfer 11 categories of Years to PPI (biological network, 50 categories of States, network average degree of 19.609) with a network average degree of 24.407 and 25.409 respectively. The social networks of Blog3 and Facebook transfers 39 categories of Interests with the network average degree of 64.776 and 43.691 respectively. The show that unsupervised CDNR works especially well in dense networks, however, domains share rare natural similarities still can't guarantee a good knowledge transfer (Blog32PPI: Interests to States). 

In addition, the general results on the Facebook dataset (Tables \ref{tab:CDNRfacebookmicro} and \ref{tab:CDNRfacebookmacro}) show promising improvements by CDNR compared to other baseline algorithms. Unsupervised representations of CDNR allow learning from small categories to large categories, and in a heterogeneous label space. CDNR uses its \textit{CD2L-RandomWalk} learning algorithm to capture the useful topologies in a large-scale information network.

\subsection{Statistical Significance}

\begin{table}[!tp]
	\centering \small
	\addtolength{\tabcolsep}{-5pt}
	\caption{Pairwise t-test results of CDNR versus baseline algorithms.}
	\label{Table:resultscore}
	\begin{tabular}{c|ccccc|ccccc}
		\hline
		& \multicolumn{5}{c|}{CDNRdblp2M10} &  \multicolumn{5}{c}{CDNRBlog32PPI}    \\ \hline
		& DeepWalk & LINE & Node2Vec & Struc2Vec & DeepGL & DeepWalk & LINE & Node2Vec & Struc2Vec & DeepGL \\ \hline
		Micro-F1 & 3.78E-13 & 7.29E-12 & 7.39E-07 & 8.13E-11 & 5.47E-07 & 3.66E-09 & 2.73E-07 & 1.03E-08 & 1.88E-08 & 8.98E-08 \\ \hline
		Macro-F1 & 1.04E-11 & 1.15E-08 & 4.31E-04 & 6.22E-10 & 5.07E-06 & 2.92E-04 & 3.17E-10 & 5.51E-03 & 7.26E-09 & 2.93E-09 \\ \hline \hline
		& \multicolumn{5}{c|}{CDNRarXivCit-HepPh2PPI} & \multicolumn{5}{c}{CDNRarXivCitHepTh2PPI} \\ \hline
		& DeepWalk & LINE & Node2Vec & Struc2Vec & DeepGL & DeepWalk & LINE & Node2Vec & Struc2Vec & DeepGL \\ \hline
		Micro-F1 & 1.38E-08 & 2.77E-07 & 1.75E-08 & 2.65E-08 & 9.47E-08 & 1.12E-07 & 1.53E-06 & 2.51E-08 & 1.92E-07 & 2.02E-06 \\ \hline
		Macro-F1 & 7.00E-04 & 6.25E-09 & 3.45E-03 & 1.79E-08 & 4.82E-09 & 7.15E-04 & 1.04E-07 & 3.18E-03 & 1.74E-07 & 4.45E-07 \\ \hline \hline
		& \multicolumn{5}{c|}{CDNRFacebook2PPI} & \multicolumn{5}{c}{CDNRBlog32Facebook} \\ \hline
		& DeepWalk & LINE & Node2Vec & Struc2Vec & DeepGL & DeepWalk & LINE & Node2Vec & Struc2Vec & DeepGL \\ \hline
		Micro-F1 & 2.20E-08 & 5.29E-07 & 1.18E-09 & 4.18E-08 & 4.61E-07 & 2.04E-03 & 9.02E-18 & 5.54E-07 & 2.95E-09 & 4.57E-03 \\ \hline
		Macro-F1 & 6.35E-04 & 7.98E-09 & 1.27E-03 & 1.25E-08 & 2.60E-08 & 7.94E-07 & 1.35E-12 & 2.65E-06 & 2.87E-09 & 9.85E-01\\ \hline
	\end{tabular}
\end{table}

To demonstrate that CDNR is indeed statistically superior to the baseline algorithms, we summarize our results for all classification evaluation tasks in Table \ref{Table:resultscore} by pairwise \textit{t}-test at a confidence level of $\alpha = 0.05$. The statistical significance is validated on every paired CDNR and baseline algorithm. On the single-label datasets, for example CDNR from the dblp dataset to the M10 dataset (CDNRdblp2M10) is compared with DeepWalk, LINE, Node2Vec, Struc2Vec and DeepGL by pairwise \textit{t}-test. 

9.02E-18 in line 11 column 8 of Table \ref{Table:resultscore} is a mean significance value averaged from nine significance values on $\{10\%,\cdots,90\%\}$ training percentages. Each of these significance values is \textit{t}-tested between CDNRBlog32Facebook and LINE. Since the CDNR multi-label dataset experiment is conducted across five datasets,the statistical significance is validated for each scenario; for example CDNRBlog32PPI is CDNR from Blog3 to PPI, and 3.66E-09 in line 3 column 7 is averaged from the nine significance values by pairwise \textit{t}-testing CDNRBlog32PPI and DeepWalk.

In Table \ref{Table:resultscore}, each value less than $\alpha = 0.05$ indicates that the difference is statistically significant. The results in Table \ref{Table:resultscore} confirm that CDNR statistically outperforms DeepWalk, LINE, Node2Vec, Struc2Vec and DeepGL in all cases expect CDNRBlog32Facebook on Macro-F1 in DeepGL which caused by the inferior results on \{80\%,90\%\} training percentages.

\subsection{Parameter Sensitivity}

\begin{figure}[!t]
	\small
	\centering
	\subfigure[dblp2M10]{\includegraphics[width=0.295\linewidth]{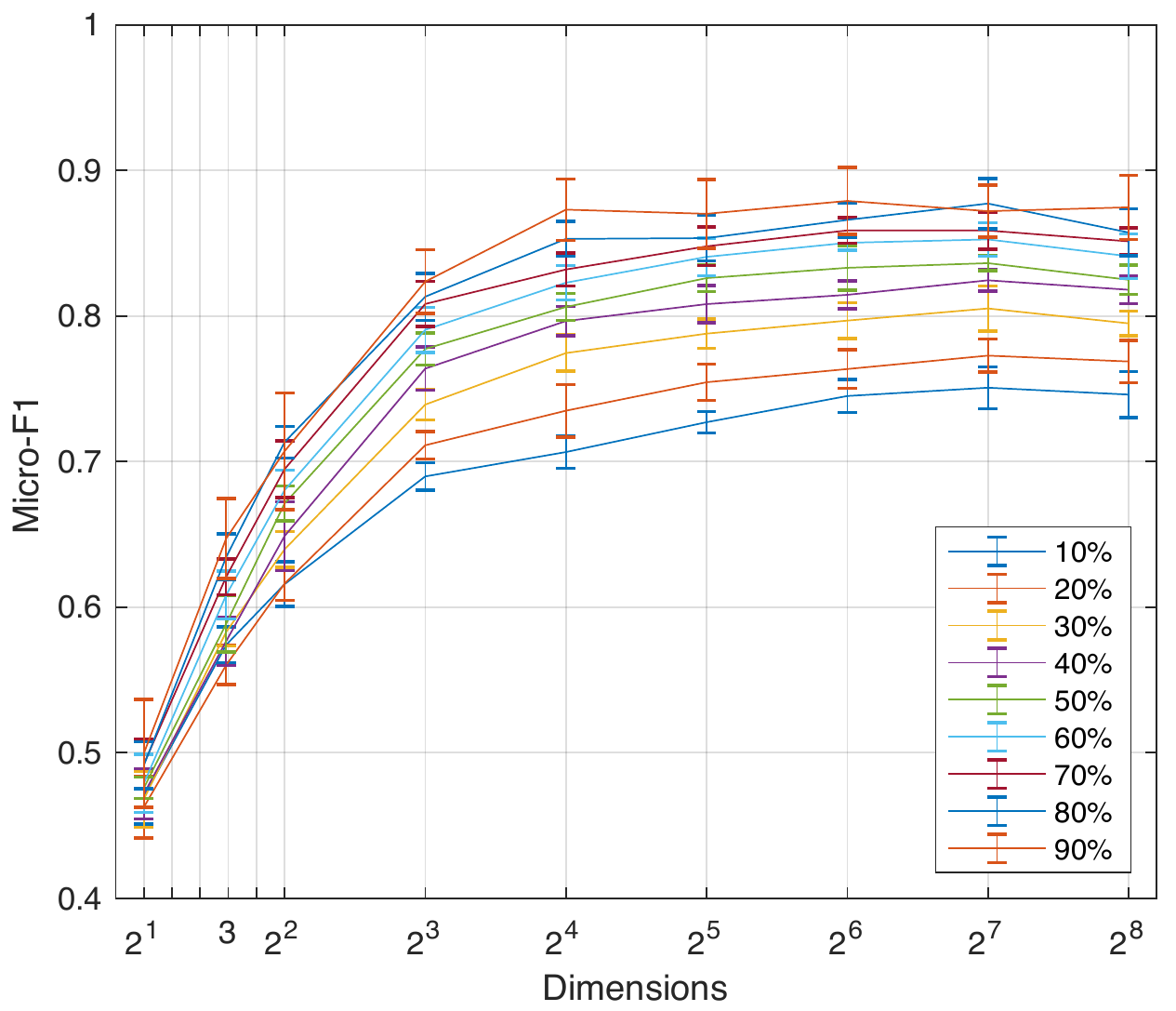}}
	\subfigure[Blog32PPI]{\includegraphics[width=0.3\linewidth]{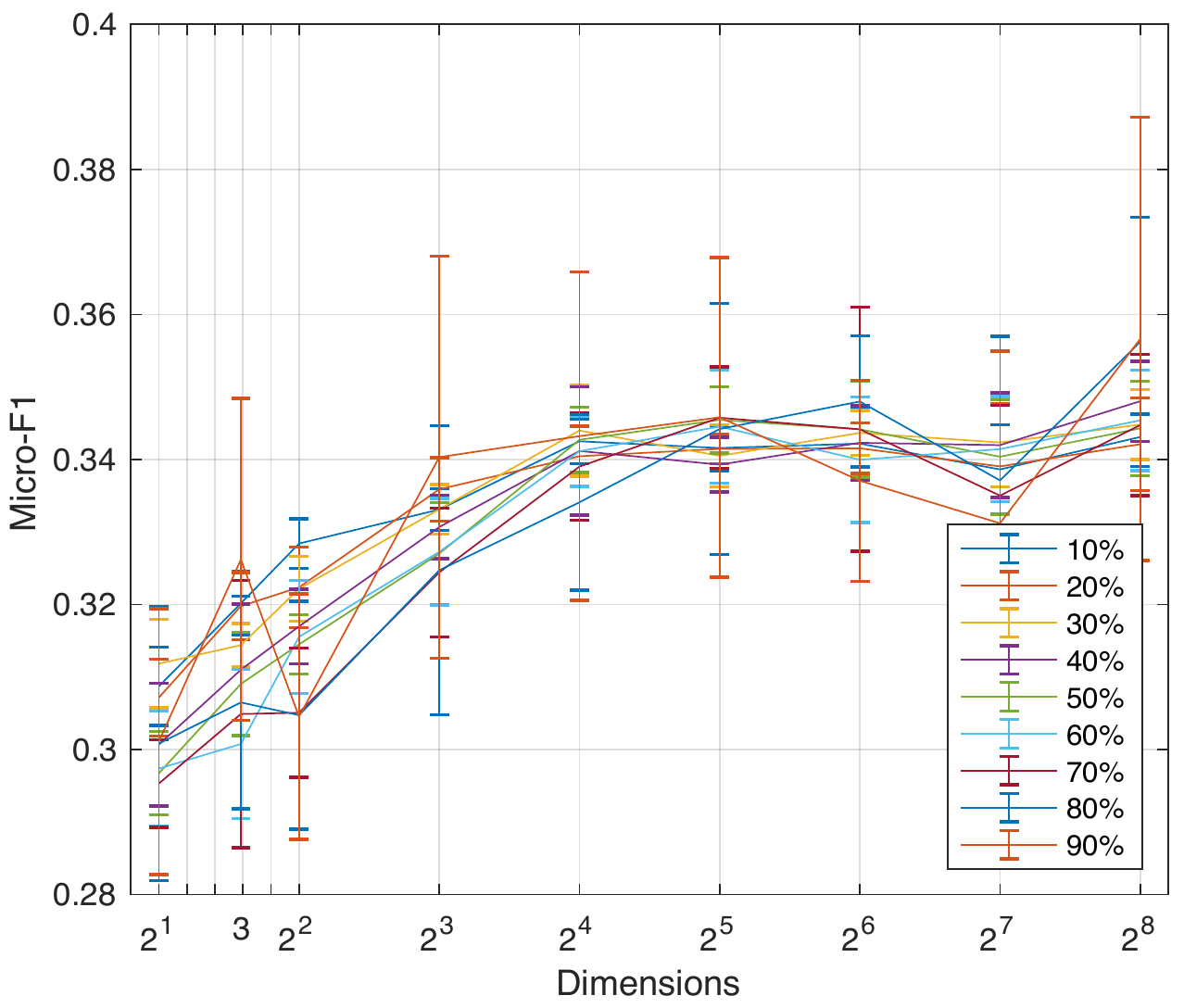}}
	\subfigure[arXivCit-HepPh2PPI]{\includegraphics[width=0.3\linewidth]{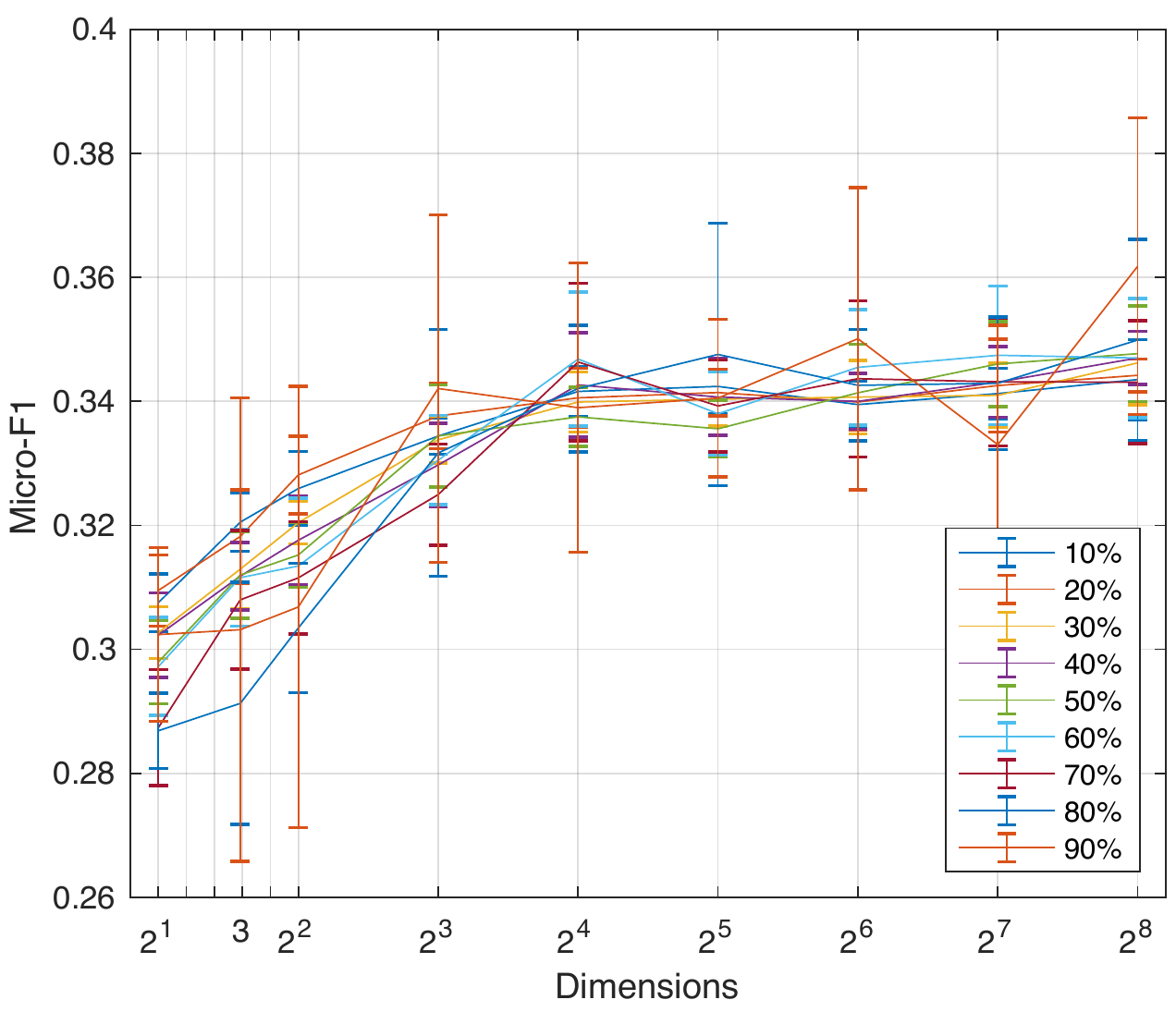}}\\
	\subfigure[arXivCit-HepTh2PPI]{\includegraphics[width=0.3\linewidth]{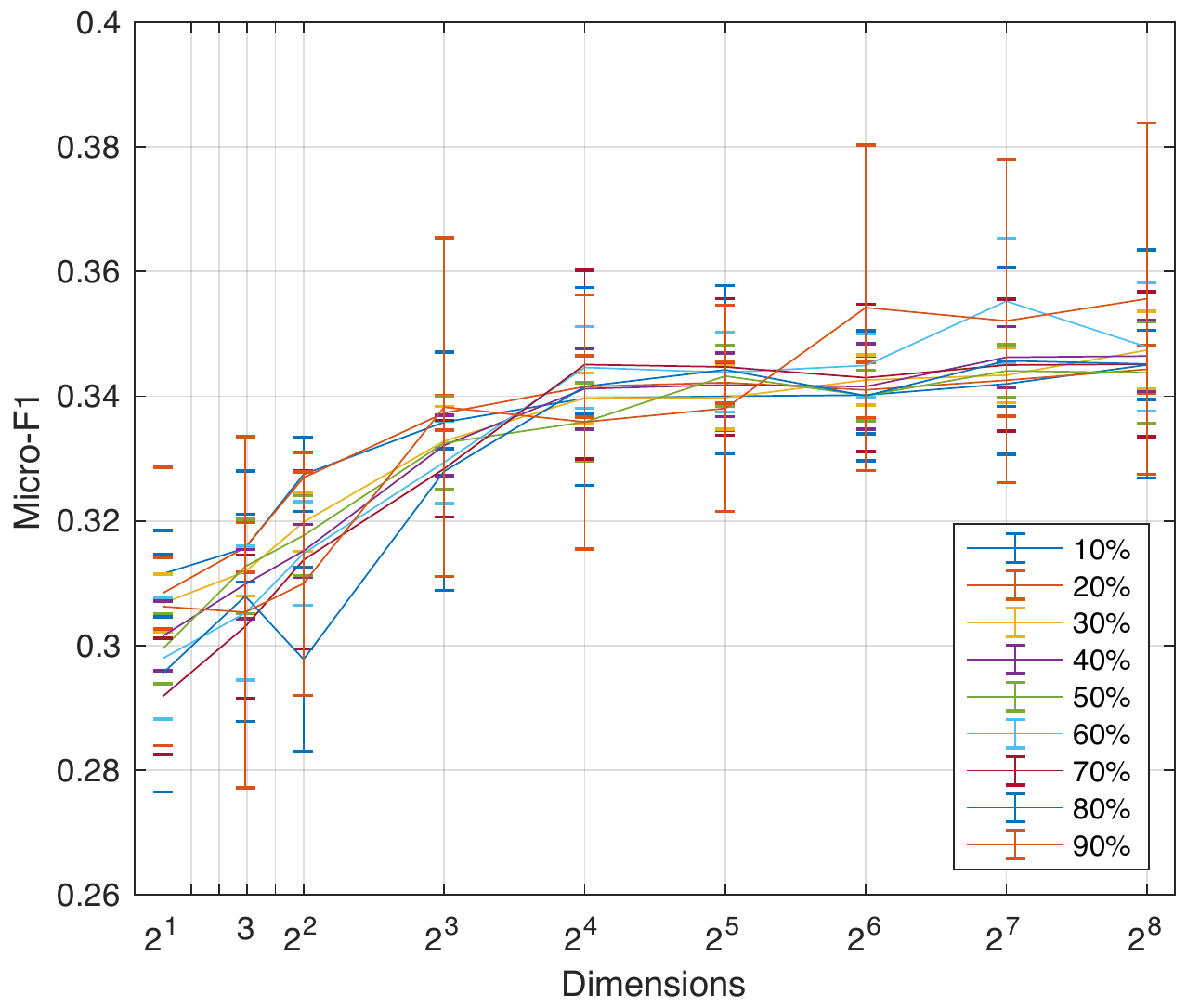}}
	\subfigure[Facebook2PPI]{\includegraphics[width=0.3\linewidth]{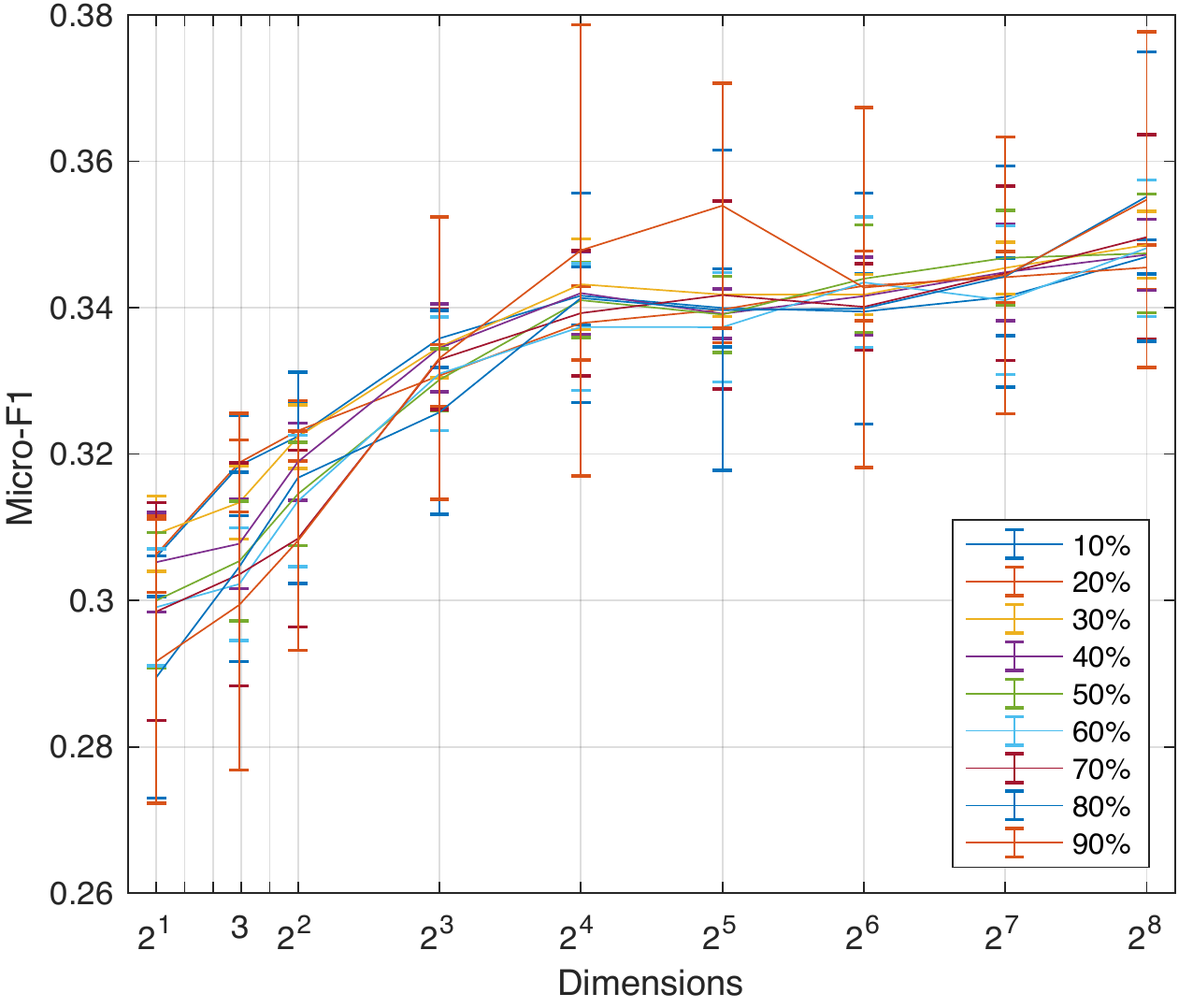}}
	\subfigure[Blog32Facebook]{\includegraphics[width=0.305\linewidth]{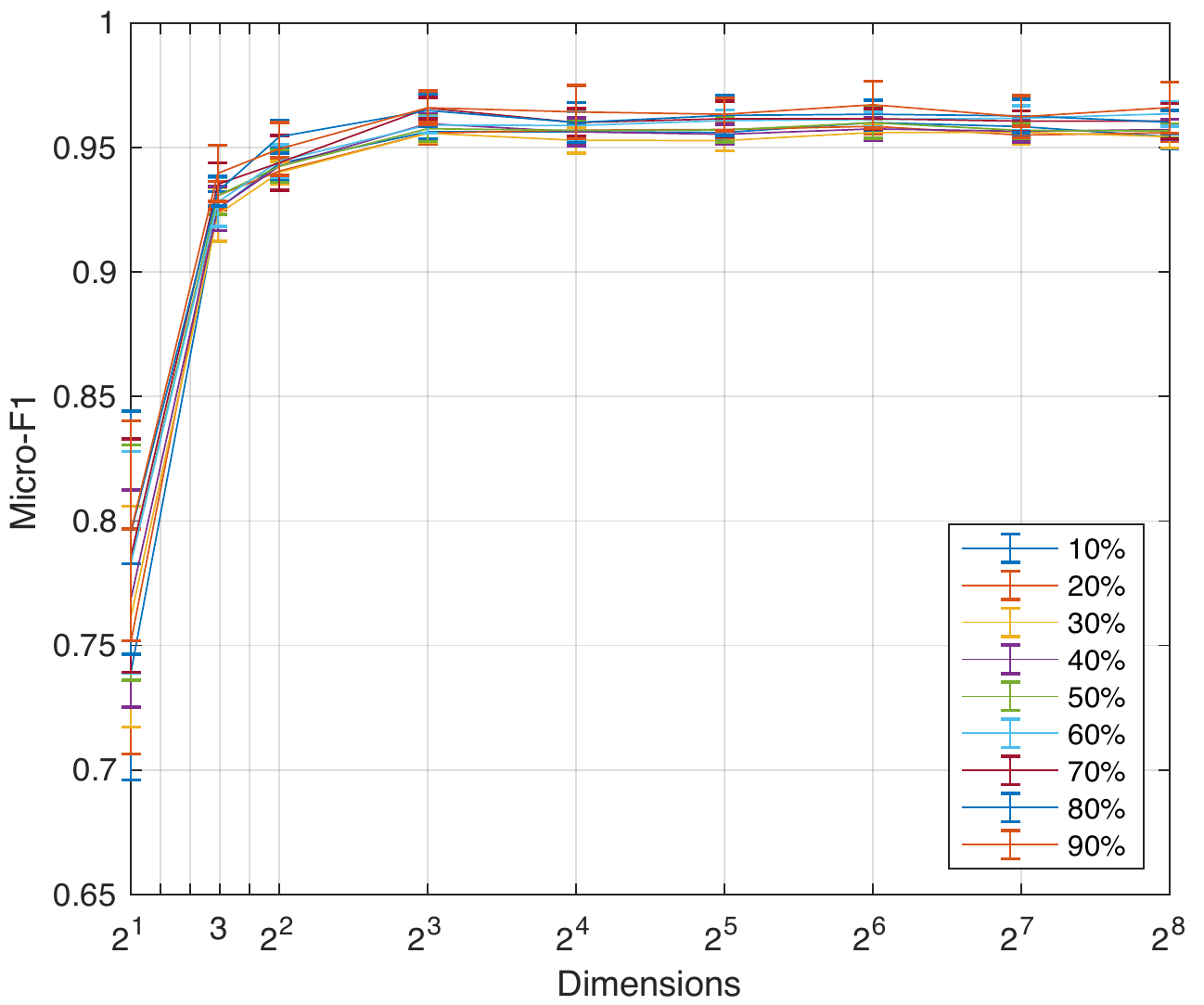}}
	\caption{CDNR dimensional sensitivity on $d=\{2, 3, 4,8,16,32,64,128,256\}$. Lines in nine colors denote dimensional sensitivity on training sample percentages \{10\%,20\%,30\%,40\%,50\%,60\%,70\%,80\%,90\%\}. The error bars in each line on $d$ points reflect the variance in 10-time testing.}
	\label{fig:parameterd}
\end{figure}

\begin{figure}[!t]
	\small
	\centering
	\subfigure[M10]{\includegraphics[width=0.3\linewidth]{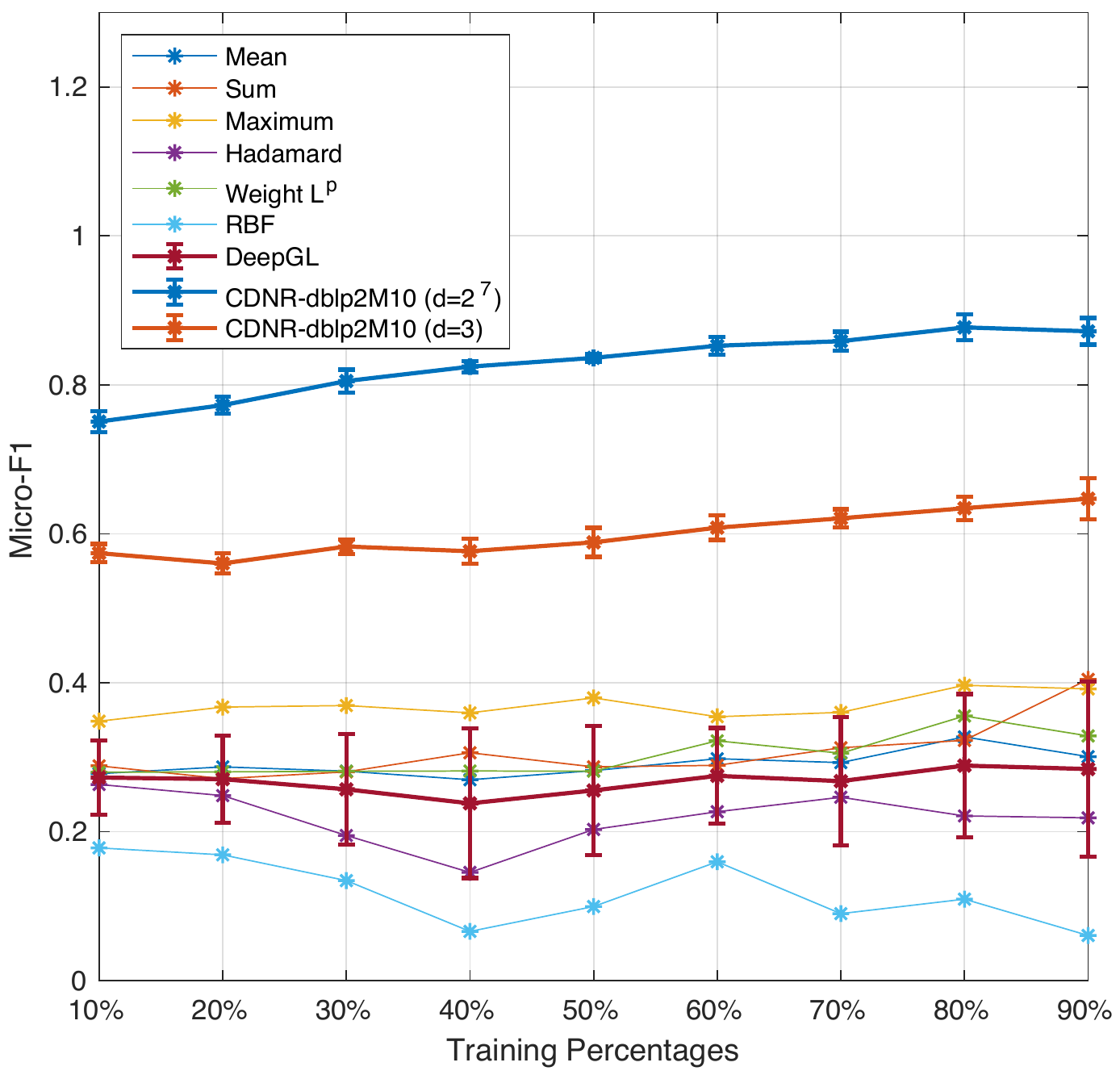}}
	\subfigure[PPI]{\includegraphics[width=0.305\linewidth]{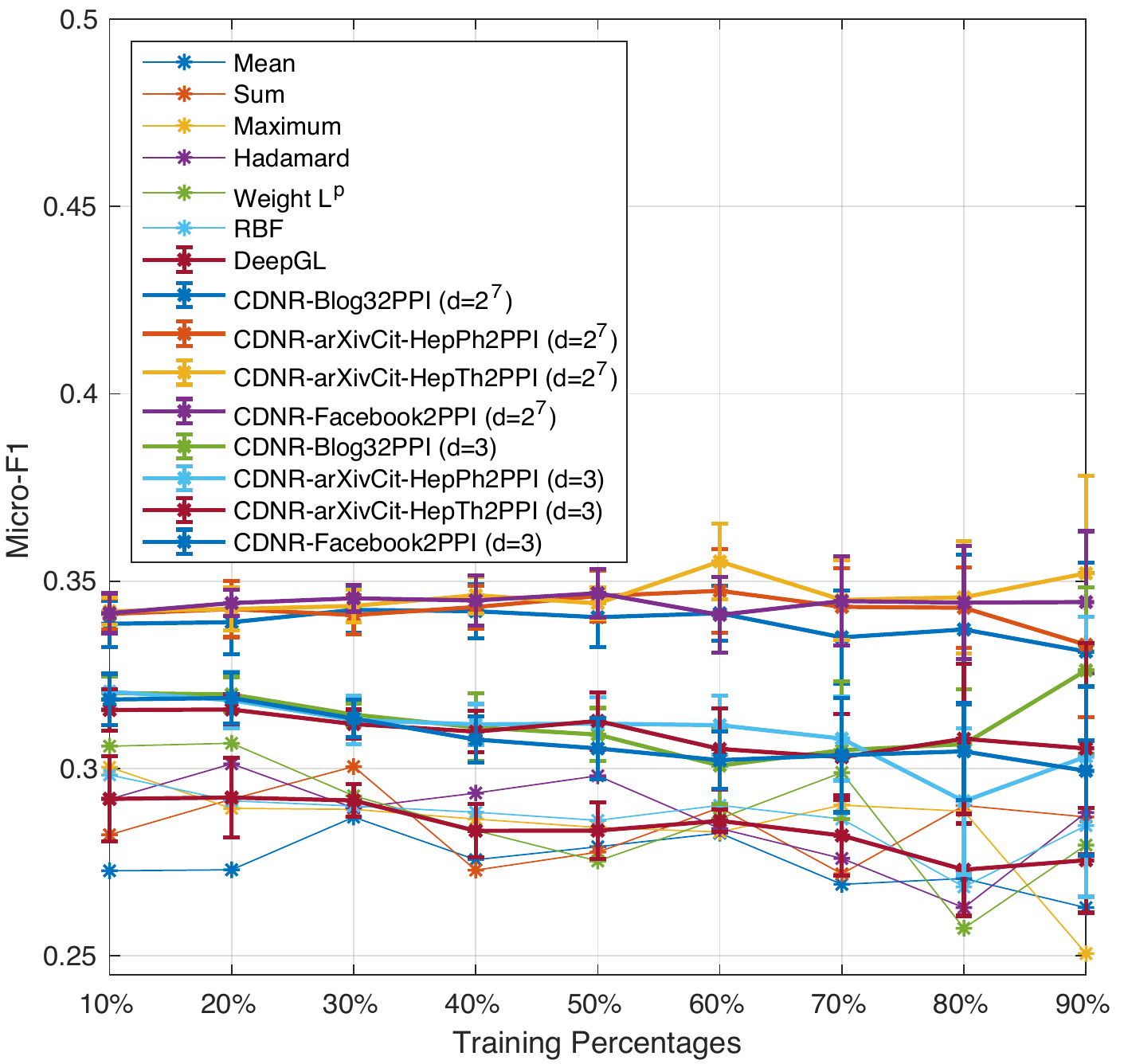}}
	\subfigure[Facebook]{\includegraphics[width=0.3\linewidth]{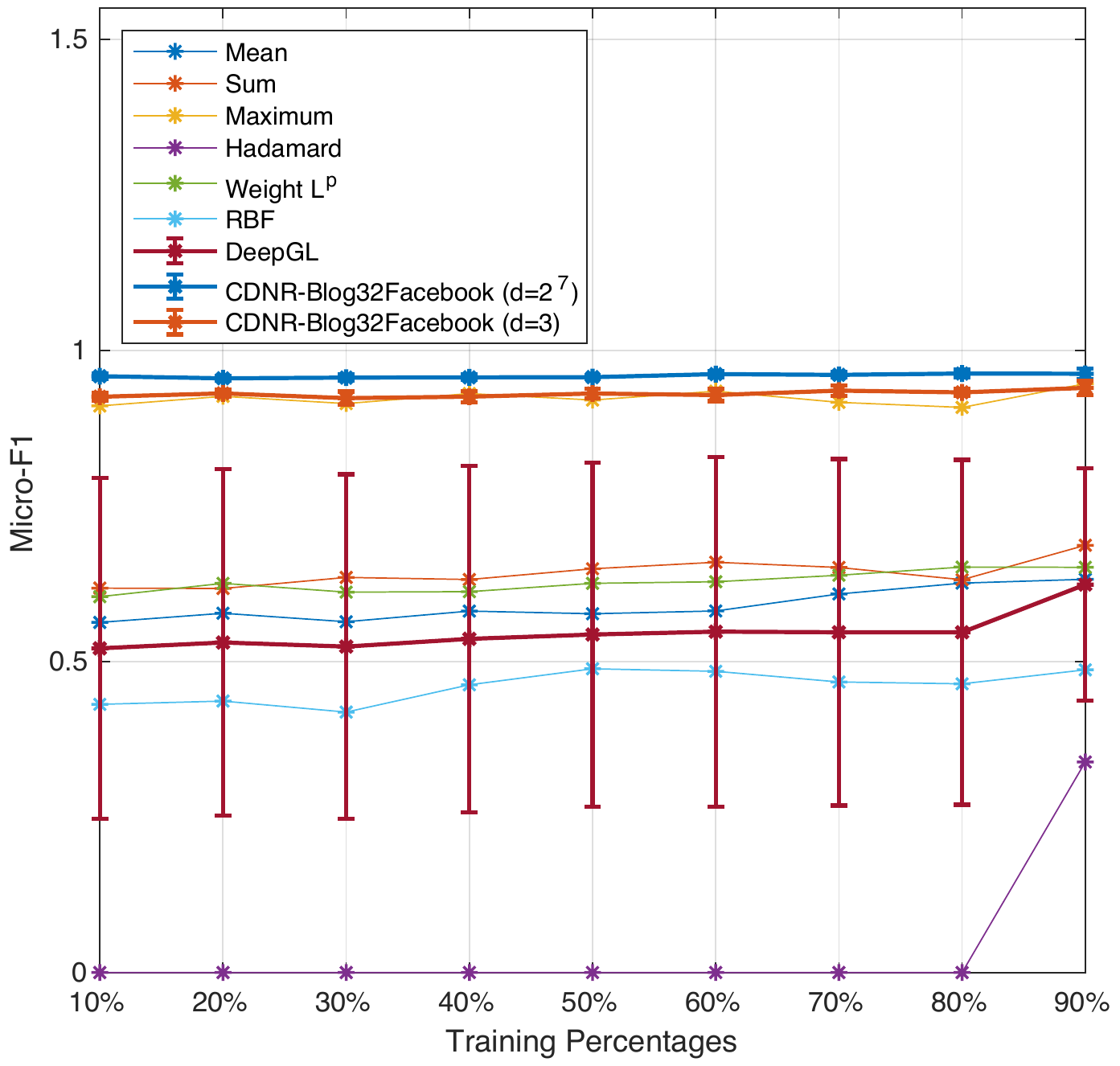}}
	\caption{Dimensional sensitivity analysis between CDNR and DeepGL-base. The thin lines denoted by operator = \{Mean, Sum, Maximum, Hadamard, Weighted $L^p$, RBF\} are tested on the 3-dimensional ($d=3$) base features generated by DeepGL; the red bold line is drawn from the average DeepGL-base experimental results; the orange bold line represents the CDNR experimental results when the dimensional parameter $d$ is set in 3; and the blue bold line represents the CDNR experimental results in $d=128$. Experiments on PPI dataset get more CDNR results by transferring knowledge from different source domains.}
	\label{fig:CDNRvsDeepGL}
\end{figure}

In this experiment, CDNR sets the representational dimension $d$ in 128 following the setting in baselines of DeepWalk, LINE, Node2Vec and Struc2Vec. However, in DeepGL, representational dimensions are determined by deep neural network training. The benefit of DeepGL, according to \cite{rossi2018deep}, is that it is able to determine the appropriate number of features automatically, as opposed to setting it to some fixed value. Unlike other techniques, DeepGL derives new feature layers as long as new and informative features are found. There is at least one new feature from the current layer remaining after pruning. In order to address the emphasis that dimensional parameter $d$ puts on network representations, we set up a parameter sensitivity testing in this part. 

CDNR on M10, PPI and Facebook setting in different feature learning dimensions, $d$=\{2, 3, 4,8,16,32,64,128,256\}, are evaluated in Figure \ref{fig:parameterd}. On average, representations trend to converge when $d>16$ and get a convergence on $d=128$. 

In Tables \ref{tab:resultsingle} and \ref{tab:CDRNPPImicro}-\ref{tab:CDNRfacebookmacro}, the representation performance of CDNR in $d=128$ has been proved significant compared with DeepGL in $d_{DeepGL}$ that varies in learning. In the format of [Minimum $d_{DeepGL}$, Average $d_{DeepGL}$, Maximum $d_{DeepGL}$], the DeepGL feature dimensions on M10, PPI and Facebook are $[57,66,80]$, $[51,61,70]$, $[56,64,82]$, respectively. In this part, the representation dimensional sensitivities in CDNR are compared with DeepGL-base. To set up, DeepGL is tested on the 3-dimensional base features; correspondingly, CDNR is evaluated on the representations in $d=3$; and the performances in $d=3$ are compared with the CDNR performance in $d=128$. The sensitivity results in Figure \ref{fig:CDNRvsDeepGL} show that CDNR on $d=128$ achieves the best performance than CDNR $d=3$ and DeepGL-base; DeepGL significantly relies on operators which represent large variance in the red bold line; and DeepGL-base can barely reach a convergence. 

In conclusion, for a common real-world network with sparse structure, CDNR outperforms DeepGL by outputting a dense feature matrix with relatively smaller dimensional size. When evaluating against DeepGL, it explains that DeepGL outputs a sparse feature matrix in contrast to other approaches even if it is larger. However, fixing the number of features is arbitrary in the case of DeepGL. Therefore, CDNR shows less dimensional sensitivity in cross-domain feature learning.

\section{Conclusions}\label{sec:Conclusions}

This work proposed a solution for a new random walk-based CDNR problem. Compared to previous network representation approaches, CDNR enables effective knowledge transfer from the external domains. Two key components flexibly tackle the challenges. The algorithm is general for universal real-world networks and is computational efficient for knowledge transferring from large-scale networks with runtime that is linear in the number of edges of the target network. CDNR has all the desired properties: flexible with any kind of networks for variety of domains and learning scenarios, effective for sampling network structures from source domain, efficient for learning from gained knowledge, and accurate with a mean improvement in F1 score of 30.68\%. Future works of similarity learning across domains and between networks based on network patterns will be studied to address the limitation in CDNR.

\section*{Acknowledgments}
The authors acknowledge the support of the Australian Research Council under DP 170101632.

\section*{References}

\bibliography{PR2018bibliography}

\end{document}